\documentclass[]{aastex631}

\shorttitle{Pre-SN neutrino search in KamLAND and Super-Kamiokande}
\shortauthors{The KamLAND and Super-Kamiokande Collaborations}

\graphicspath{{./}
}

\begin{document}

\newcommand{\solarmass}{\,M$_{\odot}$~}
\newcommand{\parsec}{\,pc~}

\title{Combined Pre-Supernova Alert System with KamLAND and Super-Kamiokande}



\newcommand{\tohoku}{\affiliation{Research Center for Neutrino Science, Tohoku University, Sendai 980-8578, Japan}}
\newcommand{\obihiro}{\affiliation{Department of Human Science, Obihiro University of Agriculture and Veterinary Medicine, Obihiro 080-8555, Japan}}
\newcommand{\osaka}{\affiliation{Graduate School of Science, Osaka University, Toyonaka, Osaka 560-0043, Japan}}   
\newcommand{\rcnp}{\affiliation{Research Center for Nuclear Physics (RCNP), Osaka University, Ibaraki, Osaka 567-0047, Japan}}
\newcommand{\tokushima}{\affiliation{Department of Physics, Tokushima University, Tokushima 770-8506, Japan}}
\newcommand{\ipmu}{\affiliation{Institute for the Physics and Mathematics  of the Universe, The University of Tokyo, Kashiwa 277-8568, Japan}}

\newcommand{\lbl}{\affiliation{Nuclear Science Division, Lawrence Berkeley National Laboratory, Berkeley, CA 94720, USA}}
\newcommand{\hawaii}{\affiliation{Department of Physics and Astronomy, University of Hawaii at Manoa, Honolulu, HI 96822, USA}}
\newcommand{\MIT}{\affiliation{Massachusetts Institute of Technology, Cambridge, MA 02139, USA}}
\newcommand{\bu}{\affiliation{Boston University, Boston, MA 02215, USA}}
\newcommand{\hdsi}{\affiliation{Halıcıoğlu Data Science Institute, University of California San Diego, La Jolla, CA, 92093}}
\newcommand{\ucsd}{\affiliation{Department of Physics, University of California San Diego, La Jolla, CA, 92093}}
\newcommand{\ut}{\affiliation{Department of Physics and Astronomy,  University of Tennessee, Knoxville, TN 37996, USA}}
\newcommand{\tunl}{\affiliation{Triangle Universities Nuclear Laboratory, Durham, North Carolina 27708, USA and Physics Departments at Duke University, North Carolina Central University, and the University of North Carolina at Chapel Hill}}    
\newcommand{\northcarolina}{\affiliation{North Carolina Central University, Durham, NC 27701, USA}}
\newcommand{\duke}{\affiliation{Department of Physics at Duke University, Durham, NC 27705, USA}}
\newcommand{\washington}{\affiliation{Center for Experimental Nuclear Physics and Astrophysics, University of Washington, Seattle, WA 98195, USA}}
\newcommand{\virginia}{\affiliation{Center for Neutrino Physics, Virginia Polytechnic Institute and State University, Blacksburg, VA 24061, USA}}
\newcommand{\nikhef}{\affiliation{Nikhef and the University of Amsterdam, Science Park, Amsterdam, The Netherlands}}
\newcommand{\udel}{\affiliation{University of Delaware, Newark, DE 19716, USA}}

\author{S.~Abe} \tohoku
\author{M.~Eizuka} \tohoku
\author{S.~Futagi} \tohoku
\author{A.~Gando} \tohoku
\author{Y.~Gando} \tohoku \obihiro
\author{S.~Goto} \tohoku
\author{T.~Hachiya} \tohoku
\author{K.~Hata} \tohoku
\author{K.~Ichimura} \tohoku
\author{S.~Ieki} \tohoku
\author{H.~Ikeda} \tohoku
\author{K.~Inoue} \tohoku
\author{K.~Ishidoshiro} \tohoku
\author{Y.~Kamei} \tohoku
\author{N.~Kawada} \tohoku
\author{Y.~Kishimoto} \tohoku 
\author{M.~Koga} \tohoku \ipmu
\author{M.~Kurasawa} \tohoku
\author{T.~Mitsui} \tohoku
\author{H.~Miyake} \tohoku
\author{D.~Morita} \tohoku
\author{T.~Nakahata} \tohoku
\author{R.~Nakajima} \tohoku
\author{K.~Nakamura} \tohoku
\author{R.~Nakamura} \tohoku
\author{R.~Nakamura} \tohoku
\author{J.~Nakane} \tohoku
\author{H.~Ozaki} \tohoku
\author{K.~Saito} \tohoku
\author{T.~Sakai} \tohoku
\author{I.~Shimizu} \tohoku
\author{J.~Shirai} \tohoku
\author{K.~Shiraishi} \tohoku
\author{R.~Shoji} \tohoku
\author{A.~Suzuki} \tohoku
\author{A.~Takeuchi} \tohoku
\author{K.~Tamae} \tohoku
\author{H.~Watanabe} \tohoku
\author{K.~Watanabe} \tohoku

\author{S.~Yoshida} \osaka
\author{S.~Umehara} \rcnp

\author{K.~Fushimi} \tokushima
\author{K.~Kotera} \tokushima
\author{Y.~Urano} \tokushima

\author{B.~E.~Berger} \lbl
\author{B.~K.~Fujikawa} \lbl \ipmu

\author{J.~G.~Learned} \hawaii
\author{J.~Maricic} \hawaii

\author{Z.~Fu} \MIT
\author{J.~Smolsky} \MIT
\author{L.~A.~Winslow} \MIT

\author{Y.~Efremenko} \ut \ipmu

\author{H.~J.~Karwowski} \tunl
\author{D.~M.~Markoff} \tunl
\author{W.~Tornow} \tunl \ipmu

\author{S.~Dell'Oro} \virginia
\author{T.~O'Donnell} \virginia

\author{J.~A.~Detwiler} \washington \ipmu
\author{S.~Enomoto} \washington \ipmu

\author{M.~P.~Decowski} \nikhef \ipmu
\author{K.~M.~Weerman} \nikhef 

\author{C.~Grant} \bu
\author{H.~Song} \bu

\author{A.~Li} \hdsi \ucsd

\author{S.~N.~Axani} \udel
\author{M.~Garcia} \udel

\collaboration{0}{The KamLAND Collaboration}
\noaffiliation

\newcommand{\AFFicrr}{\affiliation{Kamioka Observatory, Institute for Cosmic Ray Research, University of Tokyo, Kamioka, Gifu 506-1205, Japan}}
\newcommand{\AFFkashiwa}{\affiliation{Research Center for Cosmic Neutrinos, Institute for Cosmic Ray Research, University of Tokyo, Kashiwa, Chiba 277-8582, Japan}}
\newcommand{\AFFipmu}{\affiliation{Kavli Institute for the Physics and
Mathematics of the Universe (WPI), The University of Tokyo Institutes for Advanced Study,
University of Tokyo, Kashiwa, Chiba 277-8583, Japan }}
\newcommand{\AFFmad}{\affiliation{Department of Theoretical Physics, University Autonoma Madrid, 28049 Madrid, Spain}}
\newcommand{\AFFubc}{\affiliation{Department of Physics and Astronomy, University of British Columbia, Vancouver, BC, V6T1Z4, Canada}}
\newcommand{\AFFbu}{\affiliation{Department of Physics, Boston University, Boston, MA 02215, USA}}
\newcommand{\AFFuci}{\affiliation{Department of Physics and Astronomy, University of California, Irvine, Irvine, CA 92697-4575, USA }}
\newcommand{\AFFcsu}{\affiliation{Department of Physics, California State University, Dominguez Hills, Carson, CA 90747, USA}}
\newcommand{\AFFcnm}{\affiliation{Institute for Universe and Elementary Particles, Chonnam National University, Gwangju 61186, Korea}}
\newcommand{\AFFduke}{\affiliation{Department of Physics, Duke University, Durham NC 27708, USA}}
\newcommand{\AFFgifu}{\affiliation{Department of Physics, Gifu University, Gifu, Gifu 501-1193, Japan}}
\newcommand{\AFFgist}{\affiliation{GIST College, Gwangju Institute of Science and Technology, Gwangju 500-712, Korea}}
\newcommand{\AFFuh}{\affiliation{Department of Physics and Astronomy, University of Hawaii, Honolulu, HI 96822, USA}}
\newcommand{\AFFicl}{\affiliation{Department of Physics, Imperial College London , London, SW7 2AZ, United Kingdom }}
\newcommand{\AFFkek}{\affiliation{High Energy Accelerator Research Organization (KEK), Tsukuba, Ibaraki 305-0801, Japan }}
\newcommand{\AFFkobe}{\affiliation{Department of Physics, Kobe University, Kobe, Hyogo 657-8501, Japan}}
\newcommand{\AFFkyoto}{\affiliation{Department of Physics, Kyoto University, Kyoto, Kyoto 606-8502, Japan}}
\newcommand{\AFFliv}{\affiliation{Department of Physics, University of Liverpool, Liverpool, L69 7ZE, United Kingdom}}
\newcommand{\AFFmiyagi}{\affiliation{Department of Physics, Miyagi University of Education, Sendai, Miyagi 980-0845, Japan}}
\newcommand{\AFFnagoya}{\affiliation{Institute for Space-Earth Environmental Research, Nagoya University, Nagoya, Aichi 464-8602, Japan}}
\newcommand{\AFFkmi}{\affiliation{Kobayashi-Maskawa Institute for the Origin of Particles and the Universe, Nagoya University, Nagoya, Aichi 464-8602, Japan}}
\newcommand{\AFFpol}{\affiliation{National Centre For Nuclear Research, 02-093 Warsaw, Poland}}
\newcommand{\AFFsuny}{\affiliation{Department of Physics and Astronomy, State University of New York at Stony Brook, NY 11794-3800, USA}}
\newcommand{\AFFokayama}{\affiliation{Department of Physics, Okayama University, Okayama, Okayama 700-8530, Japan }}
\newcommand{\AFFosaka}{\affiliation{Department of Physics, Osaka University, Toyonaka, Osaka 560-0043, Japan}}
\newcommand{\AFFox}{\affiliation{Department of Physics, Oxford University, Oxford, OX1 3PU, United Kingdom}}
\newcommand{\AFFqmul}{\affiliation{School of Physics and Astronomy, Queen Mary University of London, London, E1 4NS, United Kingdom}}
\newcommand{\AFFregina}{\affiliation{Department of Physics, University of Regina, 3737 Wascana Parkway, Regina, SK, S4SOA2, Canada}}
\newcommand{\AFFseoul}{\affiliation{Department of Physics, Seoul National University, Seoul 151-742, Korea}}
\newcommand{\AFFsheff}{\affiliation{Department of Physics and Astronomy, University of Sheffield, S3 7RH, Sheffield, United Kingdom}}
\newcommand{\AFFshizuokasc}{\affiliation{Department of Informatics in Social Welfare, Shizuoka University of Welfare, Yaizu, Shizuoka, 425-8611, Japan}}
\newcommand{\AFFstfc}{\affiliation{STFC, Rutherford Appleton Laboratory, Harwell Oxford, and Daresbury Laboratory, Warrington, OX11 0QX, United Kingdom}}
\newcommand{\AFFskk}{\affiliation{Department of Physics, Sungkyunkwan University, Suwon 440-746, Korea}}
\newcommand{\AFFtodai}{\affiliation{Department of Physics, University of Tokyo, Bunkyo, Tokyo 113-0033, Japan }}
\newcommand{\AFFtit}{\affiliation{Department of Physics,Tokyo Institute of Technology, Meguro, Tokyo 152-8551, Japan }}
\newcommand{\AFFtus}{\affiliation{Department of Physics, Faculty of Science and Technology, Tokyo University of Science, Noda, Chiba 278-8510, Japan }}
\newcommand{\AFFtoronto}{\affiliation{Department of Physics, University of Toronto, ON, M5S 1A7, Canada }}
\newcommand{\AFFtriumf}{\affiliation{TRIUMF, 4004 Wesbrook Mall, Vancouver, BC, V6T2A3, Canada }}
\newcommand{\AFFtokai}{\affiliation{Department of Physics, Tokai University, Hiratsuka, Kanagawa 259-1292, Japan}}
\newcommand{\AFFtsinghua}{\affiliation{Department of Engineering Physics, Tsinghua University, Beijing, 100084, China}}
\newcommand{\AFFynu}{\affiliation{Department of Physics, Yokohama National University, Yokohama, Kanagawa, 240-8501, Japan}}
\newcommand{\AFFllr}{\affiliation{Ecole Polytechnique, IN2P3-CNRS, Laboratoire Leprince-Ringuet, F-91120 Palaiseau, France }}
\newcommand{\AFFbari}{\affiliation{ Dipartimento Interuniversitario di Fisica, INFN Sezione di Bari and Universit\`a e Politecnico di Bari, I-70125, Bari, Italy}}
\newcommand{\AFFnapoli}{\affiliation{Dipartimento di Fisica, INFN Sezione di Napoli and Universit\`a di Napoli, I-80126, Napoli, Italy}}
\newcommand{\AFFroma}{\affiliation{INFN Sezione di Roma and Universit\`a di Roma ``La Sapienza'', I-00185, Roma, Italy}}
\newcommand{\AFFpadova}{\affiliation{Dipartimento di Fisica, INFN Sezione di Padova and Universit\`a di Padova, I-35131, Padova, Italy}}
\newcommand{\AFFkeio}{\affiliation{Department of Physics, Keio University, Yokohama, Kanagawa, 223-8522, Japan}}
\newcommand{\AFFwinnipeg}{\affiliation{Department of Physics, University of Winnipeg, MB R3J 3L8, Canada }}
\newcommand{\AFFkcl}{\affiliation{Department of Physics, King's College London, London, WC2R 2LS, UK }}
\newcommand{\AFFwarwick}{\affiliation{Department of Physics, University of Warwick, Coventry, CV4 7AL, UK }}
\newcommand{\AFFral}{\affiliation{Rutherford Appleton Laboratory, Harwell, Oxford, OX11 0QX, UK }}
\newcommand{\AFFwu}{\affiliation{Faculty of Physics, University of Warsaw, Warsaw, 02-093, Poland }}
\newcommand{\AFFbcit}{\affiliation{Department of Physics, British Columbia Institute of Technology, Burnaby, BC, V5G 3H2, Canada }}
\newcommand{\AFFtohoku}{\affiliation{Department of Physics, Faculty of Science, Tohoku University, Sendai, Miyagi, 980-8578, Japan }}
\newcommand{\AFFicise}{\affiliation{Institute For Interdisciplinary Research in Science and Education, ICISE, Quy Nhon, 55121, Vietnam }}
\newcommand{\AFFilance}{\affiliation{ILANCE, CNRS - University of Tokyo International Research Laboratory, Kashiwa, Chiba 277-8582, Japan}}
\newcommand{\AFFibs}{\affiliation{Center for Underground Physics, Institute for Basic Science (IBS), Daejeon, 34126, Korea}}
\newcommand{\AFFglasgow}{\affiliation{School of Physics and Astronomy, University of Glasgow, Glasgow, Scotland, G12 8QQ, United Kingdom}}
\newcommand{\AFFoecu}{\affiliation{Media Communication Center, Osaka Electro-Communication University, Neyagawa, Osaka, 572-8530, Japan}}
\newcommand{\AFFminn}{\affiliation{School of Physics and Astronomy, University of Minnesota, Minneapolis, MN  55455, USA}}
\newcommand{\AFFsilesia}{\affiliation{August Che\l{}kowski Institute of Physics, University of Silesia in Katowice, 75 Pu\l{}ku Piechoty 1, 41-500 Chorz\'{o}w, Poland}}

\AFFicrr
\AFFkashiwa
\AFFmad
\AFFbu
\AFFbcit
\AFFuci
\AFFcsu
\AFFcnm
\AFFduke
\AFFllr
\AFFgifu
\AFFgist
\AFFglasgow
\AFFuh
\AFFibs
\AFFicise
\AFFicl
\AFFbari
\AFFnapoli
\AFFpadova
\AFFroma
\AFFilance
\AFFkeio
\AFFkek
\AFFkcl
\AFFkobe
\AFFkyoto
\AFFliv
\AFFminn
\AFFmiyagi
\AFFnagoya
\AFFkmi
\AFFpol
\AFFsuny
\AFFokayama
\AFFoecu
\AFFox
\AFFral
\AFFseoul
\AFFsheff
\AFFshizuokasc
\AFFsilesia
\AFFstfc
\AFFskk
\AFFtohoku
\AFFtokai
\AFFtodai
\AFFipmu
\AFFtit
\AFFtus
\AFFtriumf
\AFFtsinghua
\AFFwu
\AFFwarwick
\AFFwinnipeg
\AFFynu

\author{K.~Abe}
\AFFicrr
\AFFipmu
\author{S.~Abe}
\AFFicrr
\author{C.~Bronner}
\AFFicrr
\author{Y.~Hayato}
\AFFicrr
\AFFipmu
\author{K.~Hiraide}
\AFFicrr
\AFFipmu
\author{K.~Hosokawa}
\AFFicrr
\author{K.~Ieki}
\author{M.~Ikeda}
\AFFicrr
\AFFipmu
\author{J.~Kameda}
\AFFicrr
\AFFipmu
\author{Y.~Kanemura}
\author{R.~Kaneshima}
\author{Y.~Kashiwagi}
\AFFicrr
% ...


\collaboration{0}{The Super-Kamiokande Collaboration}
\noaffiliation

\correspondingauthor{Z. Hu, K. Saito, \& L. N. Machado} \email{hu.zhuojun.67f@st.kyoto-u.ac.jp, saito@awa.tohoku.ac.jp, lucas.nascimentomachado@glasgow.ac.uk}

\begin{abstract}
Preceding a core-collapse supernova, various processes produce an increasing amount of neutrinos of all flavors characterized by mounting energies from the interior of massive stars.
Among them, the electron antineutrinos are potentially detectable by terrestrial neutrino experiments such as KamLAND and Super-Kamiokande via inverse beta decay interactions.
Once these pre-supernova neutrinos are observed, an early warning of the upcoming core-collapse supernova can be provided.
In light of this, KamLAND and Super-Kamiokande, both located in the Kamioka mine in Japan, have been monitoring pre-supernova neutrinos since 2015 and 2021, respectively.
Recently, we performed a joint study between KamLAND and Super-Kamiokande on pre-supernova neutrino detection.
A pre-supernova alert system combining the KamLAND detector and the Super-Kamiokande detector was developed and put into operation, which can provide a supernova alert to the astrophysics community.
Fully leveraging the complementary properties of these two detectors, the combined alert is expected to resolve a pre-supernova neutrino signal from a 15\solarmass star within 510\parsec of the Earth, at a significance level corresponding to a false alarm rate of no more than 1~per century.
For a Betelgeuse-like model with optimistic parameters, it can provide early warnings up to 12~hours in advance.

\end{abstract}


\section{Introduction} \label{sec:intro}

Neutrinos emitted by a supernova during the first $\sim$10~seconds carry unique information about the physics of supernovae, which hold immense significance in the realm of astrophysics. 
The first observed supernova neutrinos~\citep{Kamiokande-II_SN, IMB_SN, Baksan_SN} were from SN1987A in the Large Magellanic Cloud, $\sim$50\,kpc away from Earth~\citep{Pietrzynski:2019cuz}. 
Since then, various neutrino experiments, such as Borexino~\citep{Borexino:2008gab}, IceCube~\citep{IceCube_SN}, KamLAND~\citep{KamLAND_SN}, LVD~\citep{LVD_SN}, NOvA~\citep{NOvA_SN}, SNO+~\citep{SNOplus_SN}, and Super-Kamiokande~(SK)~\citep{Super-Kamiokande_SN}, equipped with advanced technology and improved capabilities, continued the quest to detect supernova neutrino bursts. 
Furthermore, a number of next-generation neutrino detectors sensitive to galactic supernova neutrinos are under construction, including DUNE~\citep{DUNE_SN}, Hyper-Kamiokande~\citep{Hyper-Kamiokande:2018ofw}, JUNO~\citep{JUNO_SN}, and KM3NeT~\citep{KM3NeT_SN}.
To catch such a fleeting event, it is desirable to be alerted well before the explosion, so that astronomers and particle physicists may prepare for the observation of supernova neutrinos and possible gravitational waves as soon as the explosion happens.

The evolution of a single star, whose initial mass is greater than 8 solar masses (M$_\odot$), to its final stages prior to the core-collapse supernova (CCSN) is characterized by nuclear burning in its core due to its high temperature and density~\citep{Woosley_stellar}.\footnote{With different physical assumptions, such as rotation, or the presence of a massive companion, the evolution can be significantly different~\citep{eldridge2022new}.}
The change of the chemical composition of a star, forming concentric shells of heavier elements along its volume, is the result of nuclear fusion of heavier elements in the core. Stars at this stage are called pre-supernova~(pre-SN) stars.
The main cooling mechanism of a pre-SN star is through neutrino emissions. 
Starting from the carbon burning stage, neutrinos are produced by pair annihilation $ e^+ e^-  \rightarrow  \bar{\nu} \nu$, producing all flavours of neutrino and antineutrino pairs~\citep{odrzywolek_preSN}. 
As the star approaches core collapse, the nuclear beta decay begins to dominate.
Nuclear processes, such as beta decay, will eventually contribute more to the neutrino emission than thermal processes as the star approaches core collapse~\citep{patton}. 

These neutrinos, referred to as pre-SN neutrinos, are potentially detectable by terrestrial detectors if the progenitor is close enough to Earth~\citep{odrzywolek2}. 
They not only signal the imminent supernova, but also provide insight into the late stages of stellar evolution of massive stars. A pre-SN neutrino detection can help unravel many uncertainties associated with stellar evolution models: the physical processes that lead to a CCSN, the shell structure formation, the isotopic composition of stars, etc. It can also provide evidence to neutrino mass ordering~\citep{kat_theorical_reviewpreSN}.

The energy of pre-SN neutrinos is of sub-MeV scale or MeV scale. 
We focus on inverse beta decay~(IBD) $ \bar{\nu}_e + p \rightarrow e^{+}+n $, which has a relatively large cross section in liquid scintillator~(LS) detectors~(e.g. KamLAND~\citep{KamLAND_detector}) and water Cherenkov detectors~(e.g. SK~\citep{Super-Kamiokande:2002weg}) in the energy range of pre-SN neutrinos.
Compared to LS detectors, whose energy threshold is typically less than 1 MeV, water Cherenkov detectors are less sensitive to low-energy neutrinos because the higher energy threshold, which is 2.5 MeV in kinetic energy as shown in Section~\ref{subsec:sksel}, limits the detection of neutron capture signals.
However, starting in 2020, the SK detector was loaded with gadolinium (Gd) to improve the neutron detection efficiency~\citep{Super-Kamiokande:2021the}.
Since 2015, KamLAND has been monitoring pre-SN neutrinos and was able to provide pre-SN alerts to the astrophysics community~\citep{kamland_preSN}. 
Later in 2021, SK has also implemented an online pre-SN alert system~\citep{lucaspaper}.
As of May 2024, no alert has been issued.
Besides IBD in LS detectors and water Cherenkov detectors, distinct detection methods in other detectors could be utilized for pre-SN neutrino detection as well.
For example, coherent neutrino-nucleus scattering in future large scale dark matter direct detection experiments, is a method complementary to IBD, because it can detect all flavors of neutrinos~\citep{Raj:2019wpy}.

In this article, we introduce a joint study between KamLAND and SK on pre-SN neutrino detection. 
This combination aims at extending the reach to potential CCSN progenitors at a greater distance and reducing the warning time of pre-SN alerts.
Additionally, we present new sensitivity results individually for KamLAND and SK.
Compared to the previous study~\citep{kamland_preSN}, KamLAND has now taken more recent pre-SN neutrino models into consideration. 
The sensitivity of SK to pre-SN neutrinos with $0.01\%$ Gd by mass is presented in~\citep{lucaspaper}.
Since 2022, the Gd concentration in SK has increased to 0.03$\%$ by mass, further enhancing its capability to identify low-energy electron antineutrinos ($\bar{\nu}_e$) via IBD.
Therefore, both experiments have reoptimized their analysis strategies according to these changes.


%
%
%
%
%
%
%
%
%
%
\section{Pre-supernova neutrino model} \label{sec:presn}
To estimate the expected signal from pre-SN $\bar{\nu}_e$ in SK and KamLAND, two models for pre-SN neutrino emission during the evolution of massive stars were used: \citep{odrzywolek, odrzywolek2} and \citep{patton}. 
Both models provide data sets for the calculation of $\bar{\nu}_e$ emission during the pre-SN stage. 
\citep{odrzywolek2} provides data sets for stars with 15\solarmass and 25\solarmass and \citep{patton} for 15\solarmass and 30\,M$_{\odot}$.

The model from \citep{odrzywolek} assumes that the entire neutrino flux comes from pair annihilation. 
For the nuclear isotopic composition of the star, the model assumes a nuclear statistical equilibrium (NSE), which is a treatment only dependent on the temperature, density, and electron fraction, making it a simple flux estimated by only post-processing an already existing stellar model.
The model from \citep{patton} includes a more complete evaluation of the neutrino flux from the pre-SN star, including contributions not only from pair annihilation, but also from plasmon decay, photoneutrino process, beta decay, and electron/positron captures. 
By using the star evolution code MESA (Modules for Experiments in Stellar Astrophysics) \citep{MESA_reference}, this model couples the isotopic evolution to the stellar evolution, giving a more robust estimation of the neutrino fluxes from nuclear weak processes.

Neutrinos undergo flavor conversion, i.e. neutrino oscillations, from the point of production to the point of detection.
To calculate the expected signal from the considered models, adiabatic neutrino oscillations in the matter of the star, and neutrino oscillations in vacuum are taken into account.
For the former, the ratio of $\bar{\nu}_e$ is changed at high Mikheyev–Smirnov–Wolfenstein resonance, which depends on the mass ordering of neutrinos~\citep{Smirnov:2003da}. 
Different transition probabilities are assumed for normal and inverted neutrino mass orderings to account for the change in ratio of electron flavour neutrinos due to the dense stellar medium and the effects of neutrino oscillations in vacuum.

\begin{figure}[htb!]
    \centering
    \includegraphics[width=0.8\textwidth]{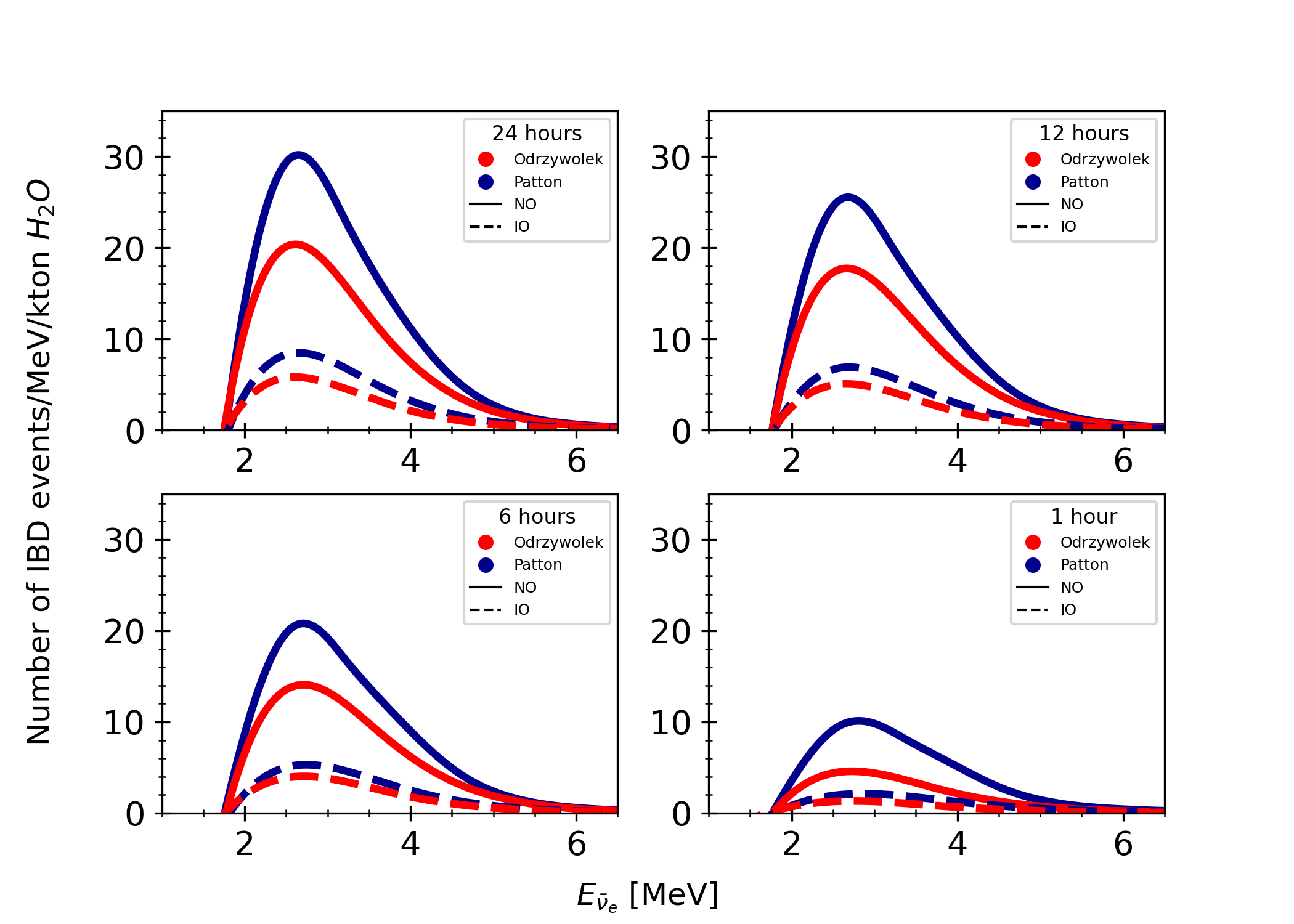}
    \caption{Number of pre-SN IBD interactions per kton of water integrated over the last 24~hours, 12~hours, 6~hours, and 1~hour prior to the CCSN as a function of the $\bar{\nu}_e$ energy, $E_{\bar{\nu}_e}$. The Betelgeuse-like models consider stars with initial masses of 15\solarmass located 150\parsec away from Earth, for both normal neutrino mass ordering (NO) and inverted neutrino mass ordering (IO).}
    \label{fig:number_IBD_nu_energy}
\end{figure}

We attempt to explore the sensitivities for detecting pre-SN neutrinos from the well-known red supergiant $\alpha$-Ori (Betelgeuse), which will potentially end up a CCSN.\footnote{There are claims that the explosion is imminent~\citep{saio2023evolutionary} and opposing views~\citep{molnar2023comment}.}
Current estimation of its mass and distance suggests 16.5-19\solarmass and $168^{+27}_{-15}$\parsec\citep{Joyce_2020}.
Limited by the available data sets, we choose 15\solarmass and 150\parsec to simulate a Betelgeuse-like pre-SN star in this work.
Although we focus on detecting pre-SN neutrinos from a Betelgeuse-like star, pre-SN neutrinos from other stars could also be observable. 
A list of candidate pre-SN stars with updated distance and mass estimates can be found in~\citep{lucaspaper}.
Figure~\ref{fig:number_IBD_nu_energy} shows the expected number of IBD candidates per kton of water for different pre-SN models as a function of the $\bar{\nu}_e$ energy, assuming a Betelgeuse-like pre-SN star.
The event spectra are obtained by integrating the expected candidates over the last 24 hours, 12 hours, 6 hours and 1 hour prior to core collapse.
These candidates predominantly cluster around $E_{\bar{\nu}_e}\approx2.6$\,MeV, but can be found at higher energies.
The expected event rate increases over time, leading to a large fraction of the total IBD candidates concentrated in the last hour.

In the following two Sections, the KamLAND and the SK experiments are introduced, accompanied by the event selection strategies.

\section{The KamLAND experiment} \label{sec:kl}
%


KamLAND is an LS detector located 1,000\,m underground in the Kamioka mine. 
KamLAND was originally designed to study reactor neutrinos,  geoneutrinos, and low-energy solar neutrinos. 
The primary target volume consists of 1\,kton of ultra-pure LS contained in a 13\,m diameter spherical balloon made of 135\,$\mu$m-thick transparent nylon ethylene vinyl alcohol copolymer (EVOH) composite film. 
The components of the KamLAND LS are 80\% dodecane and 20\% pseudocumene (1,2,4-trimethylbenzene) with 1.36\,g/L of the fluor PPO (2,5-diphenyloxazole). 
An array of 1,325 17-inch photomultiplier tubes (PMTs) and 554 20-inch PMTs mounted on the inner surface of an 18\,m diameter stainless steel sphere is used to detect the scintillation light from events occurring within the balloon. 
Non-scintillating mineral oil fills the space between the balloon and the inner surface of the sphere. 
This is all surrounded by a 3.2\,kton water Cherenkov detector contained in a resin-coated cylindrical rock cavern for cosmic-ray veto.
Detailed information of the detector is given in \citep{KamLAND_detector}.




KamLAND started its data taking in March 2002. 
Pre-SN $\bar{\nu}_e$ are expected to be detected through IBD processes, which is the main interaction channel for these neutrinos in KamLAND.
Positrons 
produced in the process lose their kinetic energy within the LS medium and annihilate with electrons, emitting two 511-keV $\gamma$ rays (prompt events). Neutrons
, with a mean lifetime of 207.5$\pm$2.8\,$\mu$s, are captured by protons, releasing 2.2\,MeV $\gamma$ rays (delayed events). 
By using the time and spatial correlation between the prompt and the delayed events, we achieve low-background conditions in the detection of $\bar{\nu}_e$.

In 2011, an inner balloon of 1.54\,m radius containing Xe-loaded liquid scintillator (Xe-LS) was installed in the center of the main balloon as a part of the KamLAND Zero-Neutrino Double-Beta Decay (KamLAND-Zen) experiment~\citep{KamLANDZen}. 
The inner balloon was updated to have a 1.92\,m radius to house double the amount of Xe-LS in 2018~\citep{KamLANDZen2022}.
The center region is not used for the $\bar{\nu}_e$ analysis because of backgrounds from the inner balloon and its support materials. 

The energy and vertex of an event can be reconstructed using the timing and charge distributions of scintillation photons recorded by the PMTs. 
The reconstruction algorithms are calibrated with radioactive sources deployed from the top of the detector~\citep{KamLAND:2009ply,Banks:2014hra}.
Using these calibration sources, the energy resolution is estimated to be ${6.4\%/\sqrt{E_{\textrm {rec}}(\textrm{MeV})}}$ and the vertex resolution is estimated to be ${12\,\text{cm}/\sqrt{E_{\textrm {rec}}(\textrm{MeV})}}$, respectively, where $E_{\textrm {rec}}$ is reconstructed energy.
The nonlinear and particle-dependent effects of the conversion between deposited (real) energy and $E_{\textrm {rec}}$ are also calibrated with the Birk’s formula~\citep{Birks:1951boa} and contribution of Cherenkov emission.



%
\subsection{Event selection in KamLAND} \label{subsec:klsel}
KamLAND performs muon vetos prior to selecting prompt-delayed pairs~(delayed coincidence method) as neutrino events. 
Cosmic-ray muons produce events with bright scintillation light and multiple spallation products, including neutrons. 
This makes it challenging to reconstruct the correct vertex and energy of low-energy events and to select prompt-delayed pairs immediately following the muon event. 
Thus, all events within 2\,ms of the arrival time of muons are vetoed. 
However, the 2-ms veto is not enough for high energy muons which make cascade showers in the detector. 
Such muons generate a non-negligible amount of long-lived spallation products such as $^{9}$Li, which has a lifetime of 257.2\,ms.
Therefore, KamLAND performs a 2-s whole-volume veto for high energy muons~\citep{KamLAND:2011bnd}. 
Alternatively, a cylindrical cut along the trajectory is applied when the reconstruction quality is good. 
These three muon vetos are determined by the total observed charge, the residual charge, which means the difference between the observed charge minus the charge that would be expected if the muon simply penetrated the detector, and the quality of muon event reconstruction. 


After applying the muon vetos, KamLAND applies the following criteria: (i) reconstructed prompt energy: ${0.9<E_p\,(\text{MeV})<4.0}$; (ii) reconstructed delayed energy: ${1.8<E_d\,(\text{MeV})<2.6}$ (capture on proton), or ${4.4 < E_d\,(\text{MeV}) < 5.6}$ (capture on $^{12}$C); (iii) spatial correlation between the prompt and delayed events: ${\Delta R < 200\,\text{cm}}$; (iv) time difference between prompt and delayed events: ${0.5 < \Delta T\,(\mu\text{s}) < 1000}$; (v) fiducial volume (FV) radii: ${R_p, R_d < 600\,\text{cm}}$; (vi) inner balloon cut:  ${R_d <2.5\,\text{m}}$ and ${\sqrt{x_d^2+y_d^2}<2.5\,\text{m}}$ for ${z_d>0\,\text{m}}$, where ${(x_d,y_d,z_d)}$ is the reconstructed delayed vertex. 
Note, the reconstructed prompt energy~($E_p$) is the sum of the positron kinetic energy and annihilation $\gamma$ energies with the quenching effect.  

Although the delayed coincidence method strongly suppresses accidental background events, KamLAND performs an additional likelihood-based selection 
to differentiate $\bar{\nu}_e$ from accidental backgrounds, which become more likely at lower energies and as the vertices are reconstructed closer to the balloon~\citep{KamLAND:2013rgu}. 

\begin{figure}[htb!]
    \centering
    \includegraphics[scale=0.4]{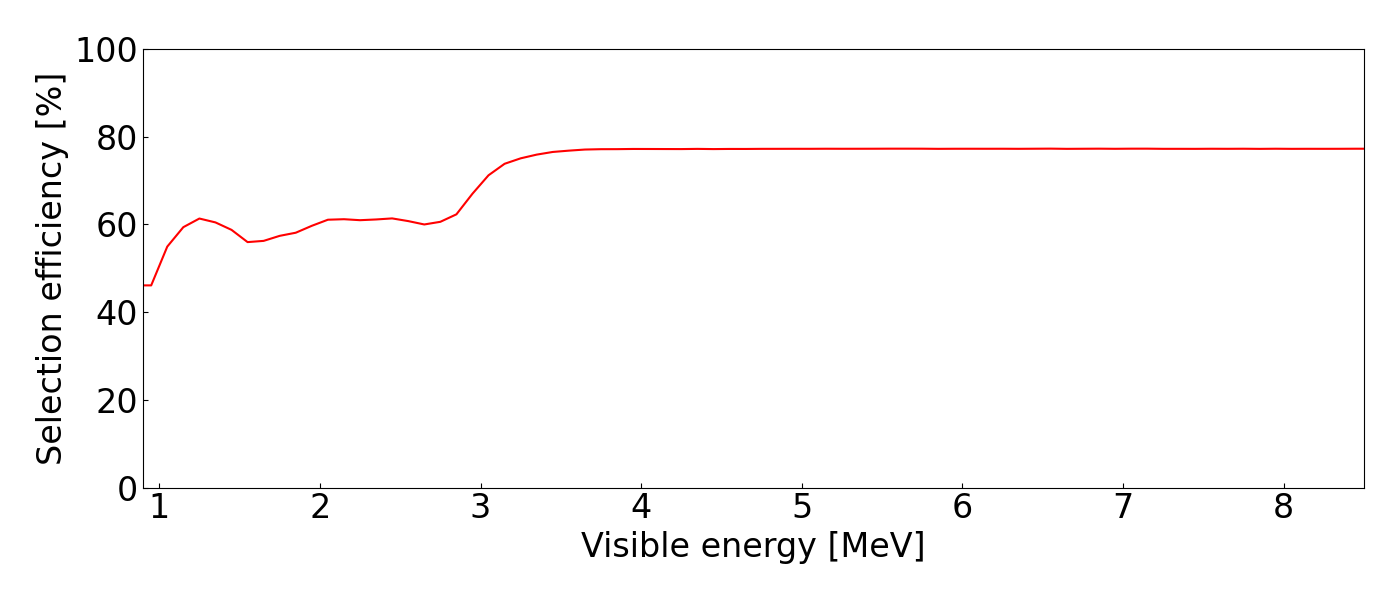}
    \caption{ The total IBD selection efficiency in KamLAND.}
    \label{fig:kam_efficiency}
\end{figure}

The total selection efficiency is calculated using a Geant4 Monte-Carlo simulation. 
A total of $10^7$ $\bar{\nu}_e$ events are generated uniformly in the 750\,cm radius volume for each prompt energy bin, and the delayed-coincidence selection with the likelihood selection are applied. 
The selection efficiency is calculated as the ratio of the number of surviving events after the selection to the number of events generated within the 600\,cm FV, shown in Figure~\ref{fig:kam_efficiency}.
The efficiency at low energies mirrors the spectrum of the accidental background because of the likelihood-based selection. 
At high energies, it remains nearly constant. 
The efficiency loss is dominated by the inner balloon cut. 
Without this cut, the efficiency is higher, $\sim$90$\%$ above 4\,MeV. 
The muon veto has an additional effect of reducing the analysis time. 
The residual analysis time after this reduction is defined as the livetime.
The KamLAND data are divided into runs.
The average lifetime ratio in any single run is approximately 0.903.

All PMT waveforms undergo digitization through front-end electronics and are collected by the Data Acquisition (DAQ) software. 
Event energy and vertex are reconstructed from these waveforms and are collected as a single file, each covering approximately 6~minutes of data. 
It takes 300--900\,s, on average 700\,s, from the time of the last event in the file to the end of the reconstruction. 
The latency is 800--1200\,s, on average 1120\,s, when we consider the first event in the file. 
The latency strongly depends on the status of other processes. 
The pre-SN monitoring process is scheduled to run at 5 minute intervals.
Upon the identification of a new file, the process applies the selection criteria described above to detect neutrino events. 
Additionally, the process counts the number of events that passed the selection criteria within the past 24 hours for pre-SN analysis.


\subsection{Background sources in KamLAND} \label{subsec:klbg}
The backgrounds for pre-SN neutrinos through IBD can be categorized into two types. One type includes non-neutrino events, such as $^{13}$C$(\alpha,n)^{16}$O reactions and accidental prompt and delayed coincidences. The other type is neutrino backgrounds, such as reactor neutrinos and geoneutrinos. 

In the early stages,  KamLAND suffered from fake prompt-delayed pairs, which are $^{13}$C$(\alpha,n)^{16}$O generated from $\alpha$-decay of $^{210}$Po in the KamLAND LS~\citep{KamLAND_alphan}. However, this $^{13}$C$(\alpha,n)^{16}$O reaction was substantially reduced during two distillation campaigns in 2007 and 2008. Currently, the rate of $^{13}$C$(\alpha,n)^{16}$O events is 0.003\,/day. The accidental background is effectively suppressed by the likelihood selection. The accidental event rate is 0.015\,/day.


Reactor $\bar{\nu}_e$ is one of the main backgrounds in this analysis and will be discussed in Section \ref{subsec:bg}.
Geoneutrinos, generated by beta decays of nuclear isotopes such as $^{238}$U and $^{232}$Th in the Earth, also constitute a background as their energies can be up to 3.27~MeV.
The expected geoneutrino event rate in KamLAND is 0.030\,/day.

\section{The Super-Kamiokande experiment} \label{sec:sk}
%

The SK experiment is a water Cherenkov detector located in the same Kamioka mine as KamLAND.
SK is a multi-purpose detector, which has been operating since 1996 and focuses on nucleon decays~\citep{Super-Kamiokande:2020wjk} and neutrino properties such as neutrino oscillations by observing atmospheric~\citep{Super-Kamiokande:2023ahc}, solar~\citep{Super-Kamiokande:2023jbt}, and accelerator neutrinos~\citep{T2K:2023smv}.
Furthermore, SK is a neutrino telescope, capable of observing neutrinos from astronomical sources~\citep{Super-Kamiokande:2021dav, Super-Kamiokande:2021sfd, Mori_2022}. 

The SK detector is composed of a cylindrical stainless steel tank with 39.3\,m diameter and 41.4\,m height~\citep{Super-Kamiokande:2002weg}. 
The detector is divided into two regions: the inner detector~(ID) and the outer detector~(OD). The ID is responsible for the event detection, with over 11,000 20-inch PMTs and it has a volume of 32\,kton, although the usual FV used in SK analyses is 22.5\,kton. 
The OD has a thickness of about 2\,m and it is composed of 1,885 8-inch PMTs, facing the outside of the detector to reduce entering cosmic-ray muon induced backgrounds.

In 2020, Gd sulfate octahydrate $\rm Gd_2(\rm SO_4)_3\cdot \rm 8H_2O$\ was dissolved to the water in the detector, starting the SK-Gd phase. 
The loading of Gd improves SK's sensitivity to low-energy $\bar{\nu}_e$, expanding the physics goals of the experiment. SK has now the potential to reveal neutrinos from the Diffuse Supernova Neutrino Background (DSNB)~\citep{Super-Kamiokande:DSNB1, Super-Kamiokande:DSNB2} and pre-SN stars, which are yet to be observed. In 2022, an additional Gd loading into SK was completed, achieving higher concentrations of Gd in the water~\citep{Super-Kamiokande:2021the}.

Low-energy $\bar{\nu}_e$ from pre-SN stars are detected in SK via IBD, similarly to KamLAND.
However, positrons generated from IBD produce Cherenkov radiation instead of scintillation light, and $\gamma$-rays from neutron capture are detected mainly by Compton scattering electrons, producing Cherenkov radiation.
In SK-Gd, the majority of thermal neutron captures occur on Gd due to its significantly higher neutron capture cross-section. Specifically, while hydrogen has a capture cross-section of only 0.3~barns, Gd's effective cross-section averages 49,000~barns.
The largest contributions for the neutron capture come from the isotopes ${}^{157}$Gd and ${}^{155}$Gd \citep{Super-Kamiokande:2021the}.
The resulting $\gamma$-ray cascade from neutron captures on Gd (nGd) releases more energy—approximately 8\,MeV—compared to captures on hydrogen, leading to a greater photon yield.
In the first phase of SK-Gd (July 2020-March 2022), which corresponded to a concentration of $0.01\%$ Gd by mass, approximately $50\%$ of neutron captures were on Gd. 
For the current phase with $0.03\%$ Gd by mass (since July 2022), the neutron capture efficiency is approximately $75\%$. 


%
\subsection{Event selection in SK} \label{subsec:sksel}

The full event selection strategy for pre-SN neutrino detection in SK is described in~\citep{lucaspaper}. Some updates have been made to the selection with the start of the second phase of SK-Gd with $0.03\%$ Gd.

The data used for the pre-SN neutrino analysis come from the Wide-band Intelligent Trigger (WIT) \citep{CARMINATI2015666}, a computing farm with approximately 900 hyper-threaded cores dedicated to real-time data processing. Each core handles 23\,ms data files sequentially, applying a set of criteria to select good-quality events while ensuring a high efficiency, even at energies as low as 2.5\,MeV in kinetic energy. After event reconstruction, the processed files are sent to an organizer machine: the files with the reconstructed events arrive time-unordered, are then organized. 
While organizing the data, the files are grouped into segments lasting about 1.5~minutes each. 
Subsequently, they are made available to the pre-SN subsystem and then transferred outside the WIT system for low-energy offline analyses. In addition to the pre-SN alert system, the WIT system also hosts an online supernova burst trigger and raw data buffer, which would be preserved in case of a supernova event. Table~\ref{tab:timing_alarm} provides the estimated time between DAQ and a decision by the pre-SN alert system.


\begin{deluxetable}{cc}[htb!]
\tablecaption{Estimated latency time of each step in the pre-SN alert system and update frequency, updated from~\citep{lucaspaper}. Total latency time is the sum of the latency of each step. }\label{tab:timing_alarm}
\tablewidth{0pt}
\tablehead{\colhead{Process} & \colhead{Estimated Time}}
\startdata
Data Fitting (WIT system) & 10 seconds \\
Data organizing (WIT system) & 4 minutes  \\
Process Queue ($\sim 2 \times 10^{6}$ events) & 2 minutes \\
Alert Decision/Export Results & Performed every 5 minutes \\
\enddata
\end{deluxetable}

\begin{figure}[htb!]
\centering
\includegraphics[width=0.6\textwidth]{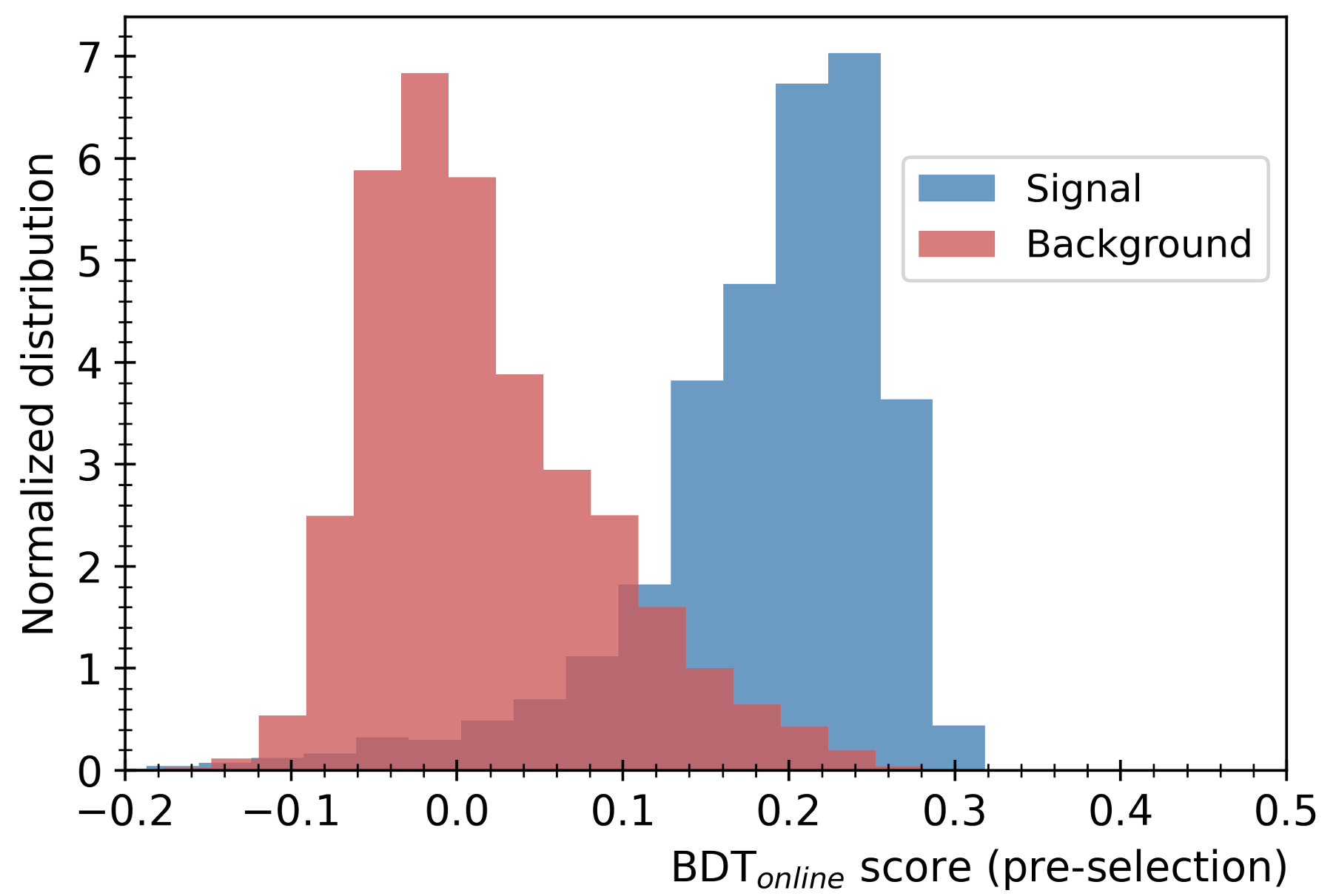}
\caption{Signal-background separation for the Boosted Decision Tree classifier used for pre-selection (BDT$_{\mathrm{online}}$) using random subsets of SK data with 0.03\% Gd as background and a fraction of the simulated IBD coincidence events as signal.}
\label{fig:bdts}
\end{figure}

\begin{figure}[htb!]
\centering
\includegraphics[width=0.6\textwidth]{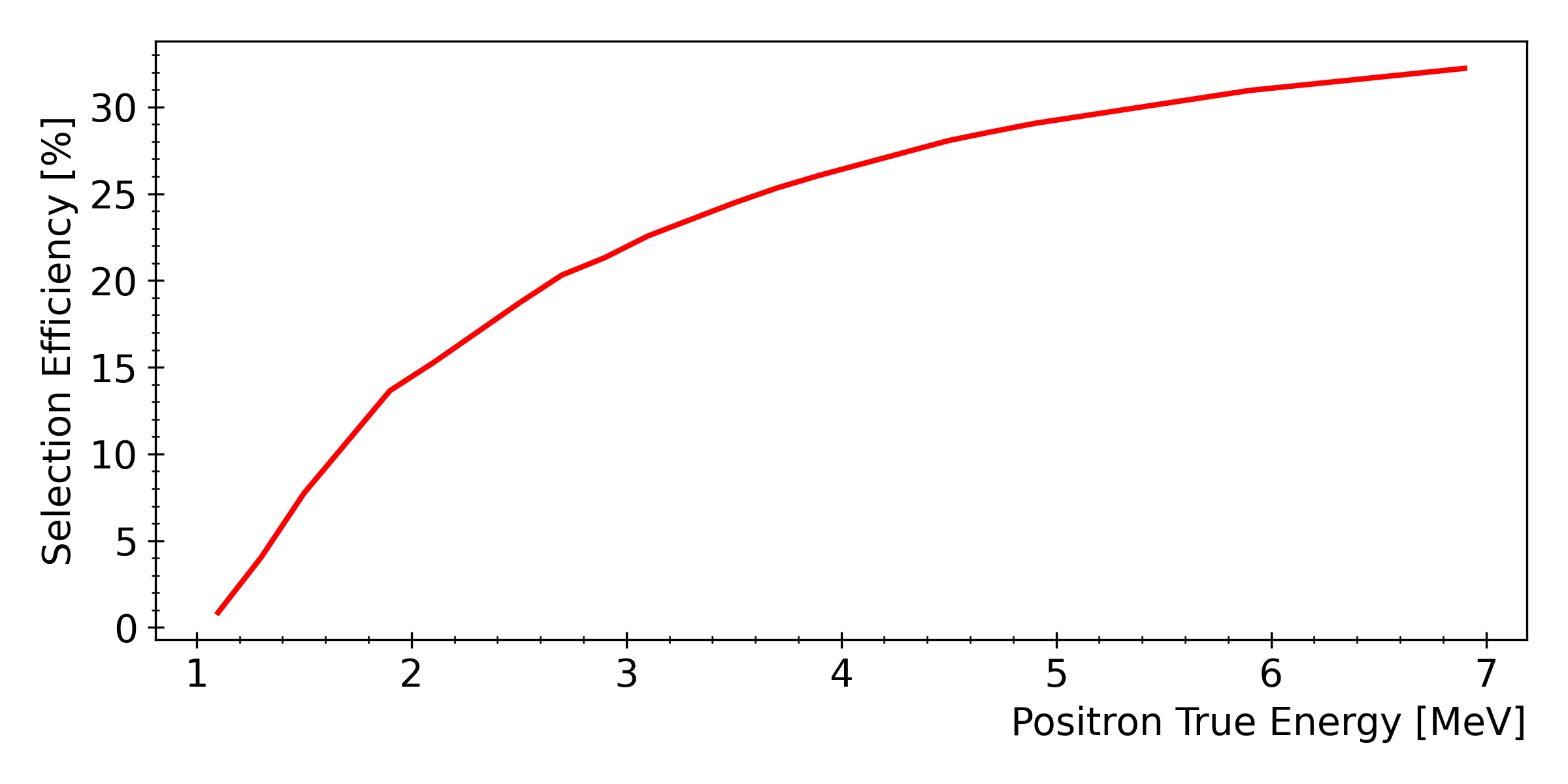}
\caption{Evaluation of the efficiency of selection after the application of cut criteria to 10$^7$ IBD pairs as a function of positron true total energy for SK.}
\label{fig:SK_selection_eff}
\end{figure}

The pre-SN alert system receives data from WIT right after the organizer processes sort the data in time. The system runs the event selection in real time, which is based on the coincidence distance ($dR$) and coincidence time ($dT$) of IBD pairs and two Boosted Decision Tree (BDT) methods: one used as pre-selection for IBD pair identification (BDT$_{\mathrm{online}}$) and another used as final selection based on angular distribution of hits, reconstructed energy and quality (BDT$_{\mathrm{offline}}$) (more details in \citep{lucaspaper}). 
For the current SK-Gd phase with $0.03\%$ Gd, BDTs were re-trained and cuts were optimized: BDT$_{\mathrm{online}} > 0.2$, $dR < 300$~cm, $dT < 80$~$\mu$s, and BDT$_{\mathrm{offline}} > -0.1$. Figure~\ref{fig:bdts} shows the signal background separation of the current BDT$_{\mathrm{online}}$ used for pre-selection and Figure ~\ref{fig:SK_selection_eff} shows the efficiency of applying the selection criteria to 10$^7$ IBD pairs.
The irreducible background rate is approximately 0.5\,event/hour.

%
\subsection{Background sources in SK} \label{subsec:skbg}

The major backgrounds for the pre-SN neutrino search in SK are reactor neutrinos and accidental coincidences. Other background sources include geoneutrinos, radioactive contaminants, and cosmic-ray muon induced spallations.
The backgrounds from geoneutrinos and accidental coincidences are similar to what has been described for KamLAND in Section~\ref{subsec:klbg}.
Reactor neutrino background will be discussed in Section~\ref{subsec:bg}.
Radioactive contaminants that came along with the Gd loading are also a background source: ${}^{235}$U chain isotopes can emit $\alpha$, contributing to the backgrounds from ${}^{18}$O($\alpha$,n)${}^{21}$Ne$^*$ and ${}^{17}$O($\alpha$,n)${}^{20}$Ne$^*$ processes.
Moreover, the spontaneous fission of ${}^{238}$U can emit neutrons that mimic delayed signals or even IBD candidates.
Cosmic-ray muon induced spallations are expected to have a tiny contribution since its resulting background rate is low, 
and are effectively removed using BDTs.
\section{Sensitivity to pre-SN neutrinos in KamLAND and SK} \label{sec:indsen}

Previous sensitivity studies for detection of pre-SN neutrinos in KamLAND and SK are given in \citep{kamland_preSN, charlespaper,lucaspaper}.
In \citep{kamland_preSN}, the sensitivity for KamLAND's detection of pre-SN neutrinos assuming the model from~\citep{odrzywolek} is presented. 
\citep{charlespaper} presents a preliminary overall sensitivity for SK doped with $0.1\%$ Gd by mass. 
In~\citep{lucaspaper}, an improved sensitivity to pre-SN neutrinos in SK is shown, using data from the first phase of SK-Gd (with $0.01\%$ Gd) to predict realistic backgrounds and new event selection methods. 

In this study, both analyses in KamLAND and SK have been updated.
KamLAND has now taken the additional pre-SN neutrino model from~\citep{patton} into consideration and reoptimized the selection parameters, the analysis time window, and the detector status. 
As for SK, it has entered a new SK-Gd phase with $0.03\%$ Gd loaded, further enhancing the sensitivity to low-energy $\bar{\nu}_e$.
The loading of Gd has brought radioactive contaminants into the FV of the detector, affecting the background rate.
Moreover, the reactor-neutrino-induced backgrounds in both detectors have changed substantially since the previous studies because many of the nuclear reactors in Japan have been restarted.
This section presents a new assessment of the sensitivity to pre-SN neutrinos in KamLAND and SK.
Analysis strategies are reoptimized to adapt to the changes.

\subsection{Analysis strategies} \label{subsec:anastg}

The general analysis strategy for each experiment is as follows. 
A rapid increase in the candidate event rate is sought without explicit reference to any of the pre-SN neutrino models.
In each experiment, the background rate is measured over a relatively long period (30 days or more) using recent data, in order to reduce the effects of random fluctuations in the data.
A sliding analysis window of a few hours is used to measure the observed event rate for the purpose of searching for signal events.
The detection significance is calculated by comparing the observed event rate to the expected background rate.
In other words, a test of significance is performed, with the null hypothesis being that the observed event rate is consistent with the expected background rate within the sampling error.

Although KamLAND and SK are at nearly identical locations, they are in rather different experimental conditions, such as target mass, detection energy threshold, background rates, and duty cycle, etc.
These factors affect the choices of background time window and analysis time window.
SK chooses a 30-day background time window, while KamLAND's choice is a longer 90-day background time window due to its lower background rate.
The analysis time windows are chosen based on the principle of achieving the longest warning time.
As a result, KamLAND has chosen an optimal time window of 24~hours.
For SK, the time window was optimized to 12~hours, maximizing the warning time for Betelgeuse-like models and reducing the impact that potential interruptions in DAQ and calibration work in the detector may have in the pre-SN alert system.

In this analysis, both experiments are considered as Poisson counting experiments, and their Poisson likelihoods $\mathcal{L}_{\mathrm{SK}}$ and $\mathcal{L}_{\mathrm{KL}}$ are constructed. 
The subscripts SK and KL denote Super-Kamiokande and KamLAND, respectively.
The Poisson likelihood for each experiment can be written as
\begin{equation}
    \mathcal{L}_{\mathrm{x}} = \frac{(\lambda_{\mathrm{x}})^{N_{\mathrm{x}}} \exp^{-\lambda_{\mathrm{x}}} }{ N_{\mathrm{x}}!},
\end{equation} 
where the subscript $\mathrm{x}$ can be SK or KL. 
$N_{\mathrm{x}}$ is the observed number of events within the sliding analysis time window.
The term $\lambda_{\mathrm{x}}$, being the expected number of events, is given by 
\begin{equation}
    \lambda_{\mathrm{x}} = S_{\mathrm{x}} + B_{\mathrm{x}},
\end{equation}
where $S_{\mathrm{x}}$ is the parameter for the number of signal and $B_{\mathrm{x}}$ is the expected number of background.
The test statistic based on the likelihood ratio is given by
\begin{equation}\label{eq:likelihoodratio}
    \Lambda_{\mathrm{x}} = -2\ln\frac{Max\left(\mathcal{L}_{\mathrm{x}}|_{S_{\mathrm{x}}=0}\right)}{Max\mathcal{\left(L_{\mathrm{x}}\right)}},
\end{equation}
where the numerator and denominator are the maximum likelihoods with and without imposing a background-only scenario $S_{\mathrm{x}}=0$~\citep{Cowan:2010js}.
The more the observation disagrees with the background-only hypothesis, the larger $\Lambda_{\mathrm{x}}$ is.
As the test statistic $\Lambda_{\mathrm{x}}$ asymptotically approaches $\chi^2$ distribution, we consider $\sqrt{\Lambda_{\mathrm{x}}}$ as the detection significance for each experiment.


\subsection{Background assumptions} \label{subsec:bg}

The background rates in KamLAND and SK can be largely affected by the nuclear reactors near the Kamioka mine. 
Reactor $\bar{\nu}_e$ are mainly generated through the beta decays of the fission products of $^{235}$U, $^{238}$U, $^{239}$Pu and $^{241}$Pu in nuclear reactors.
The Great East Japan Earthquake of 2011 caused numerous reactors in Japan to be shut down, and they have been gradually restarting since 2015. A constant monitoring of the situation in Japan's nuclear power plants is being conducted, and backgrounds are being updated.
Nonetheless, it is difficult to predict when and which reactors will start operation in the future.
Therefore, we assume different reactor background conditions according to three scenarios: low, medium and high reactor activities.
The low reactor activity scenario assumes that all reactors in Japan are not in operation.
For the medium reactor activity scenario, reactors near the Kamioka mine (Mihama 3 unit, Ohi 3,\,4 units and Takahama 1,\,2,\,3,\,4 units) are assumed to be operating with a 100\% load factor.
We note this is close to the situation as of the winter of 2023-2024.
The high reactor activity scenario assumes the amount of reactor neutrinos is doubled compared to the medium reactor activity scenario.

\begin{figure}[htb!]
\centering
\includegraphics[width=0.6\textwidth]{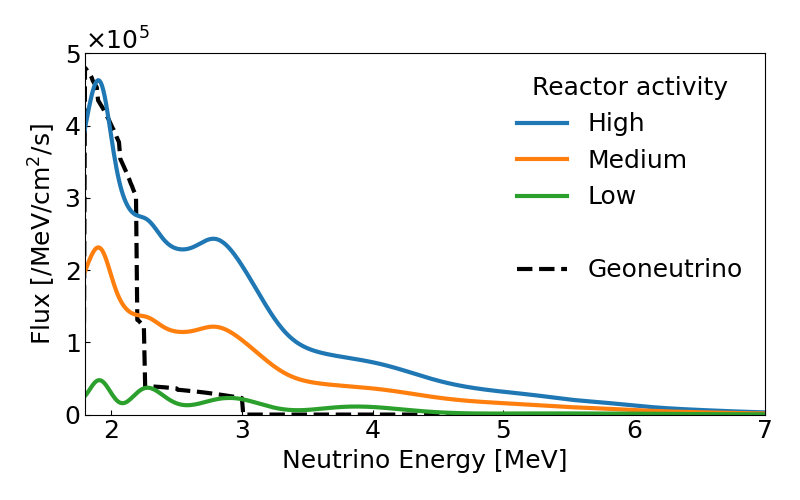}
\caption{Reactor $\bar{\nu}_e$ fluxes at the Kamioka mine assuming low, medium and high reactor activities. 
The relative fission yields ($^{235}$U,$^{238}$U,$^{239}$Pu,$^{241}$Pu) are assumed be (0.570,0.078,0.295,0.057)~\citep{KamLAND:2002uet}. 
The $\bar{\nu}_e$ spectra per fission are from \citep{Huber2011,Mueller2011,Vogel1981}.
The values of neutrino oscillation parameters are $\Delta m_{21}^2=7.53\times 10^{-5}$ eV$^2$, $\tan^2 \theta_{12}=0.436$, and $\sin^2\theta_{13}=0.023$.
Geoneutrino flux (black) is calculated using the parameters from~\citep{Enomoto_geo}.}
\label{fig:reactor_flux}
\end{figure}

Figure~\ref{fig:reactor_flux} shows the expected reactor fluxes at the Kamioka mine considering these three reactor scenarios, as well as the expected geoneutrino fluxes.
The fluxes are calculated based on the $\bar{\nu}_e$ spectra per fission from~\citep{Huber2011,Mueller2011,Vogel1981}, with relative fission yields ($^{235}$U,$^{238}$U,$^{239}$Pu,$^{241}$Pu) assumed be (0.570,0.078,0.295,0.057)~\citep{KamLAND:2002uet}.
The values of neutrino oscillation parameters are $\Delta m_{21}^2=7.53\times 10^{-5}$\,eV$^2$, $\tan^2 \theta_{12}=0.436$, and $\sin^2\theta_{13}=0.023$.
The thermal power of each reactor is estimated from publicly available data on reactor electric power.
In the low reactor activity scenario, where all reactors in Japan are assumed to be off, the only contribution we consider comes from reactors in Korea.
In addition to the reactor neutrino fluxes, we also plot the geoneutrino flux at the Kamioka mine for comparison.
The geoneutrino flux is calculated based on the model in~\citep{Enomoto_geo}.
For $\bar{\nu}_e$ energy below 2.2\,MeV, the geoneutrino flux is comparable to the reactor neutrino flux under the high reactor activity assumption.
It decreases rapidly at $\sim$2.2\,MeV, and is roughly the same as the reactor neutrino flux assuming low reactor activity between 2.2\,MeV and 3\,MeV.
Above 3\,MeV, the geoneutrino flux becomes negligible.

\subsection{False alarm rate} \label{subsec:far}

It is a common practice to quantify the false positive rate of a statistical test using the $p$-value of the test.
However, in this search, we find it misleading to report the results using the $p$-value or the significance of a single test. 
The online search of pre-SN neutrinos is performed continuously, while the time when the pre-SN neutrino signal appears is not known in advance.
An appropriate way to estimate the $p$-value is to determine the probability, assuming background-only, of finding a signal at any time that is at least as extreme as the one observed.
Such a $p$-value can be substantially higher than the $p$-value of a single test.
This is the so-called ``look-elsewhere effect"~\citep{Lyons:2008hdc}.

To resolve this issue, we introduce the quantity ``false alarm rate" to report the result of the search.
The false alarm rate is the expected frequency that a false alarm may happen per century.
A false alarm is caused by a significant signal due to background fluctuations. 
The false alarm rate can be derived from toy Monte-Carlo simulations, assuming a background-only scenario. 
In practice, we generate a time series of Poisson random events with the expected value being the background rate.
The statistical test described in the previous subsection is performed. 
Then we evaluate the frequency with which the random events cause a significant signal. 
Thereby, a false alarm rate is found.
Considering the frequency of CCSN in the galaxy is approximately once every few decades~\citep{fpr_reference}, we set a ${\text{false alarm rate} \leq 1}~\text{per century}$ as the ultimate pre-SN alert criteria.

%
\subsection{Sensitivity to pre-SN neutrinos at KamLAND} \label{subsec:klsen}
\begin{figure}[t]
\centering
\gridline{\fig{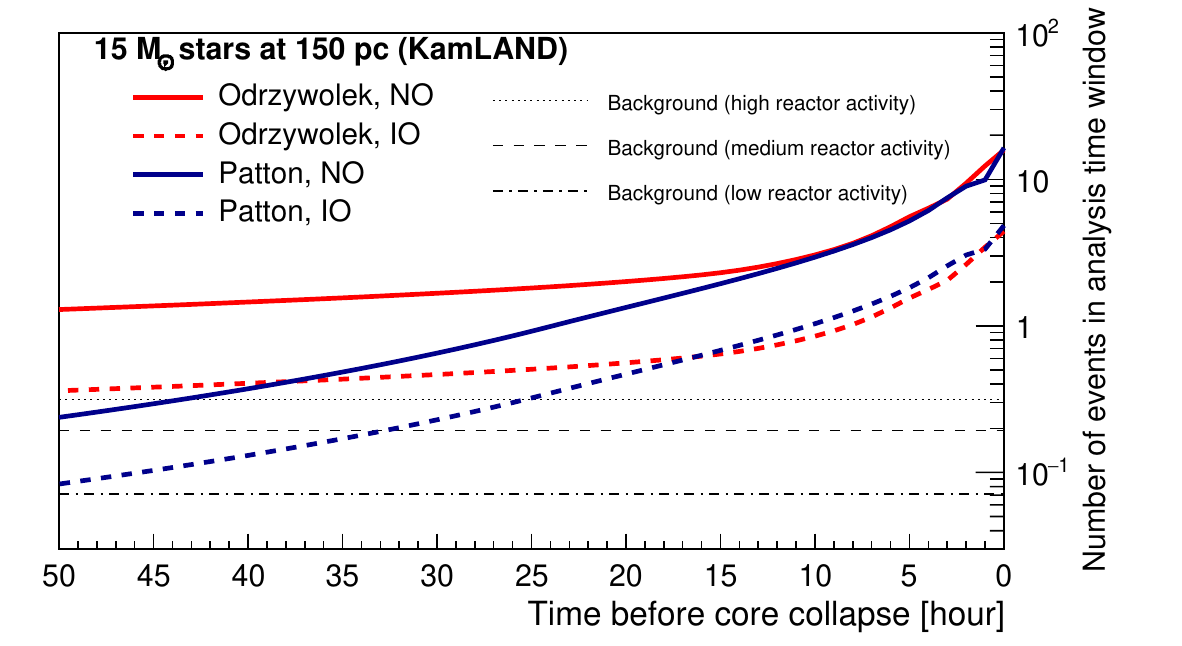}{0.48\textwidth}{(a)}
          \fig{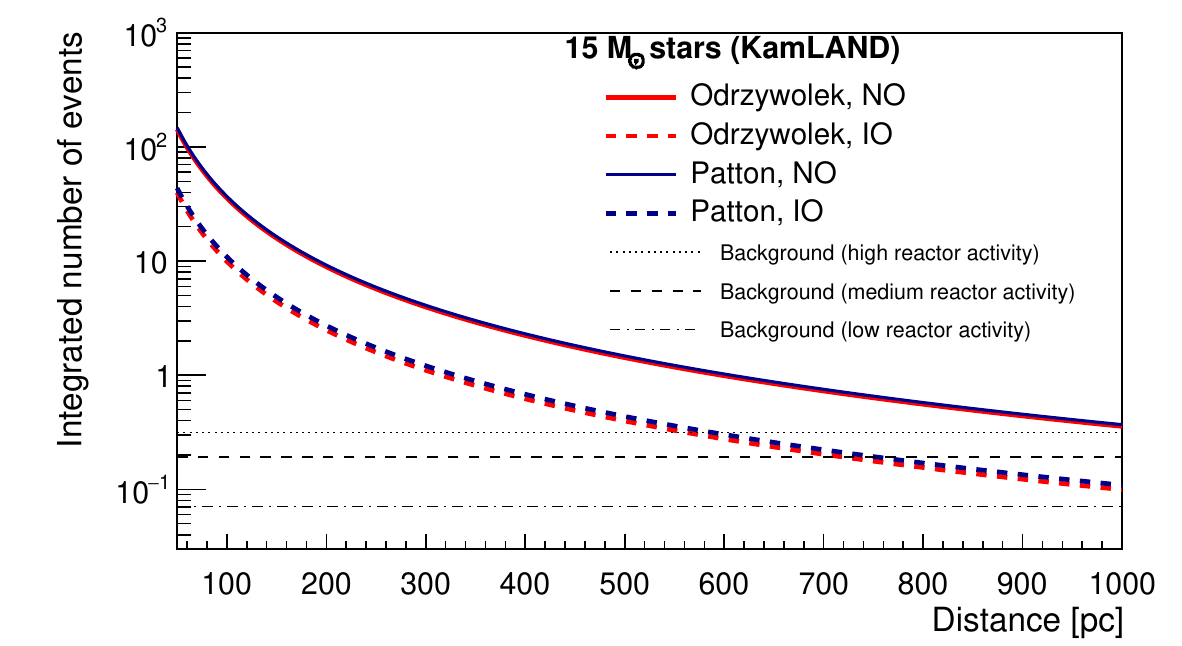}{0.48\textwidth}{(b)}
         }
\caption{\label{fig:num_kl} Expected number of signal events in KamLAND as a function of (a) time to core collapse, and (b) distance. Pre-SN $\bar{\nu}_e$ fluxes from a star with 15\solarmass are considered, following the Odrzywolek model (red) and the Patton model (blue). For (b), the signal rates are integrated over the last 24~hours before the CCSN. Solid curves show normal neutrino mass ordering and dashed curves show inverted neutrino mass ordering. }
\end{figure}

We evaluate the expected numbers of signal events in KamLAND for the two pre-SN neutrino models with neutrino oscillation effects, assuming a Betelgeuse-like pre-SN star.
Figure~\ref{fig:num_kl}~(a) shows the expected number of signal events, integrated over a sliding 24-hour time window, as a function of time to CCSN. 
Figure~\ref{fig:num_kl}~(b) shows the integrated number of signals for the last 24~hours before core collapse at different distances.
The expected background counts integrated over 24~hours are ${0.07,\,0.19,~\text{and}~0.32~\text{events}}$ for low, medium, and high reactor activities, respectively. 
The background rate in KamLAND is sufficiently low, such that a few candidate events can cause a significant signal.

\begin{figure}[t]
\centering
\includegraphics[width=0.66\textwidth]{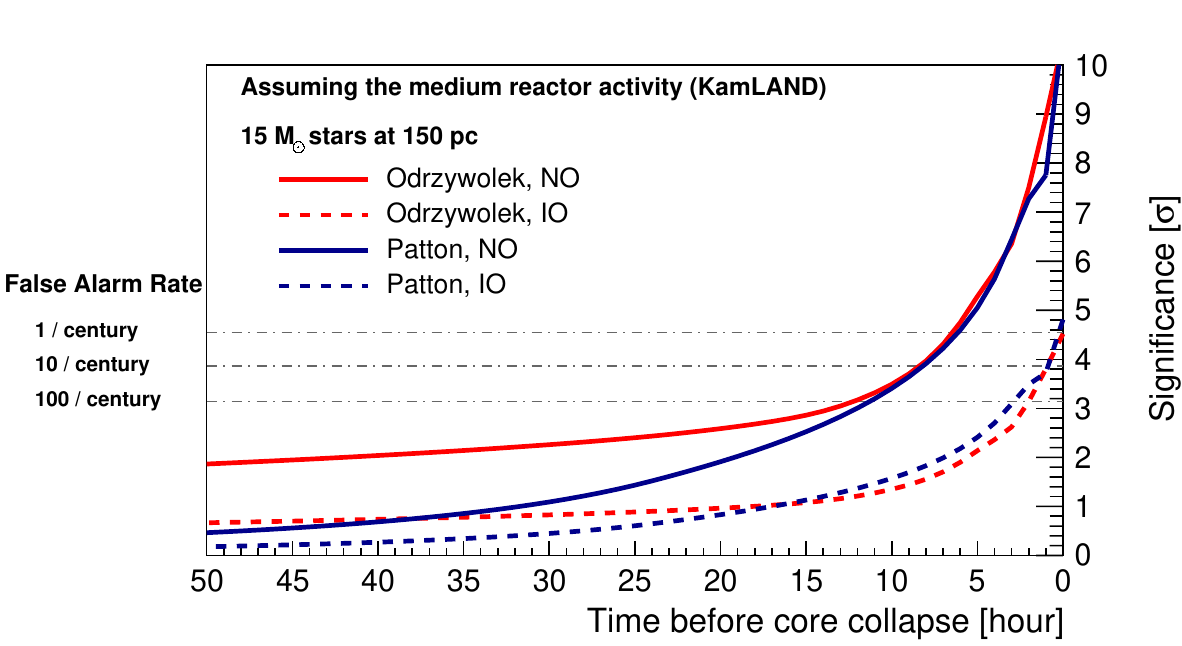}
\caption{\label{fig:sig_kl} Time evolution of the sensitivity to pre-SN neutrinos in KamLAND, assuming medium reactor activity, following the Odrzywolek model (red) and the Patton model (blue). Solid (dashed) lines are for normal (inverted) neutrino mass ordering. Horizontal dotted-dashed lines indicate false alarm rates of 1, 10, and 100~per century.}

\includegraphics[width=0.6\textwidth]{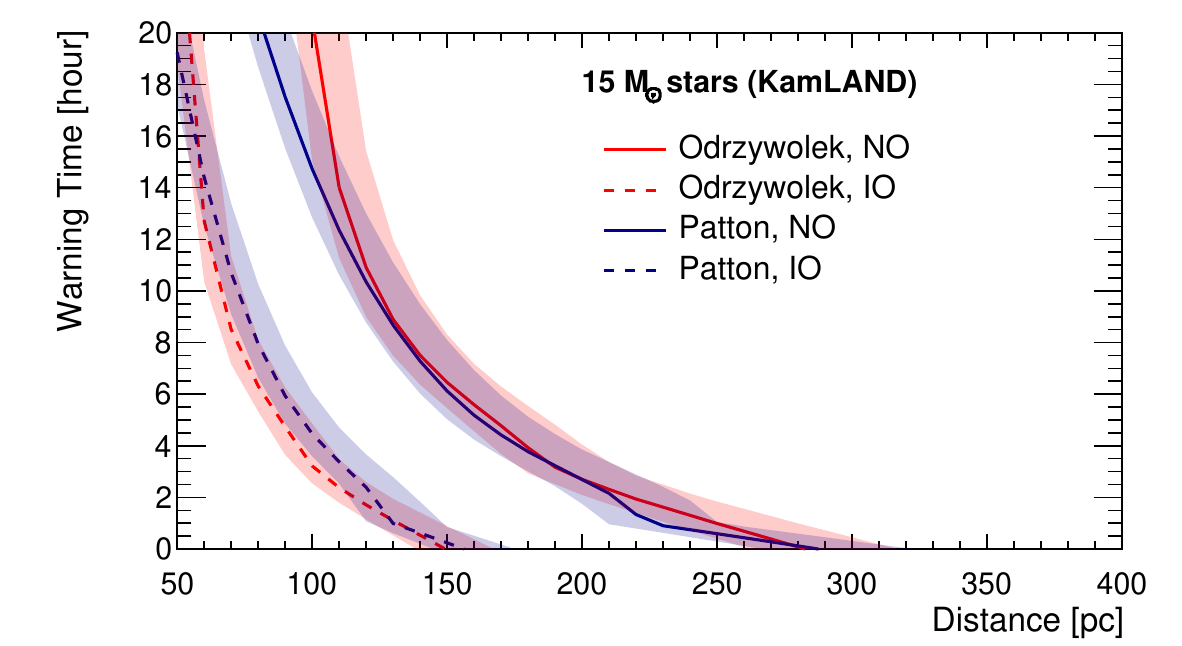}
\caption{\label{fig:dis_kl}Expected warning time in KamLAND as a function of distance. The lines are estimations assuming medium reactor activity. The upper (lower) edges of the bands are for the low (high) reactor activity case. }
\end{figure}

Figure~\ref{fig:sig_kl} presents the time evolution of the expected detection significance assuming medium reactor activity.
The significance corresponding to false alarm rates of 1, 10, and 100~per century are also plotted as horizontal dotted-dashed lines.
The remaining time to core collapse, when KamLAND observes an excess of pre-SN neutrino candidates as extreme as ${\text{false alarm rate} \leq 1}~\text{per century}$, is defined as the warning time.
Note that the latency due to data processing is not taken into account when calculating the warning time.
For a Betelgeuse-like pre-SN star, KamLAND is capable of issuing a pre-SN alert 6.5~hours before the CCSN, assuming the Odrzywolek model and normal mass ordering.
In the case of inverted mass ordering, the warning time is largely shortened as the pre-SN neutrino fluxes become lower.
The worst case is, with the Odrzywolek model and inverted mass ordering, the expected detection significance cannot reach the alert criteria.
The discussions above are based on the medium reactor activity scenario.
Table~\ref{tab:cmb_time} summarizes the warning time for all three reactor activity assumptions.
If the reactor activity assumption shifts from medium to high, for the normal ordering cases, the warning time will be shortened by roughly 1~hour.
For the inverted ordering cases, with the high reactor activity assumption, KamLAND is unable to issue an alert with a ${\text{false alarm rate of}~1~\text{per century}}$, noted as ``N/A~(Not Applicable)" in Table~\ref{tab:cmb_time}.

Figure~\ref{fig:dis_kl} pictures the warning time as a function of distance.
The lines are estimations assuming medium reactor activity.
The upper edges of the bands are for low reactor activity, and the lower edges are for high reactor activity.
These results indicate that, for nearby pre-SN candidates, KamLAND can send alerts tens of hours before the explosion.
For the medium reactor activity case, KamLAND is sensitive to pre-SN candidates within an optimistic distance of 280\parsec away from Earth.

\subsection{Sensitivity to pre-SN neutrinos at SK} \label{subsec:sksen}
%

\begin{figure}
\centering
\gridline{\fig{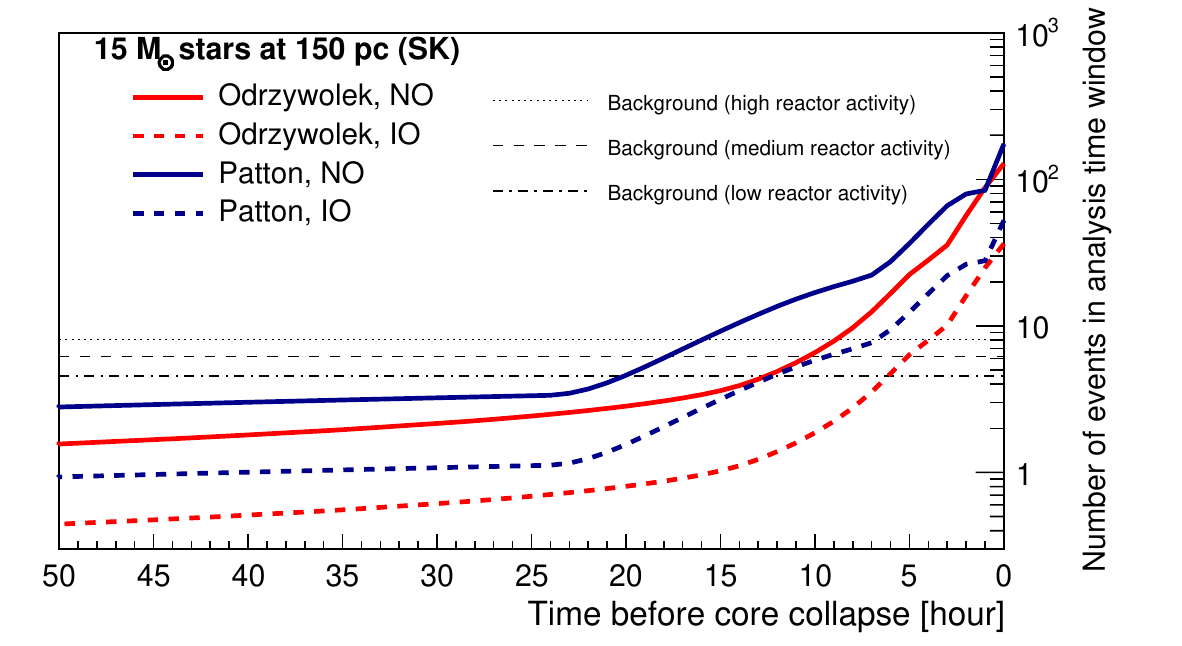}{0.48\textwidth}{(a)}
          \fig{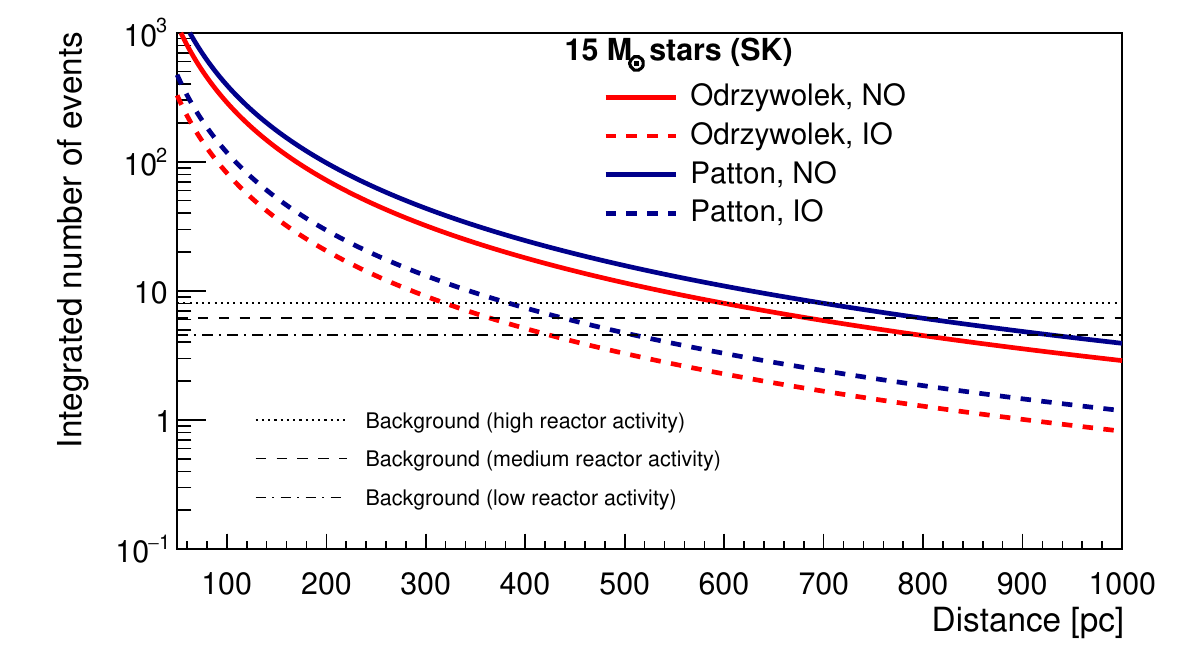}{0.48\textwidth}{(b)}
         }
\caption{\label{fig:num_sk} Expected number of signal events in SK with 0.03$\%$ Gd concentration as a function of (a) time to core collapse, and (b) distance. Pre-SN $\bar{\nu}_e$ fluxes from a star with 15\solarmass is considered, following the Odrzywolek model (red) and the Patton model (blue). For (b), the signal rates are integrated over the last 12~hours before the CCSN. Solid curves show normal neutrino mass ordering and dashed curves show inverted neutrino mass ordering. }
\end{figure}

The fluxes of pre-SN neutrinos are taken from the two pre-SN models, with neutrino oscillation effects assuming normal and inverted mass orderings.
The expected signal rates are aggregated over a sliding 12-hour time window, resulting in the expected number of signal events as a function of time, as pictured in Figure~\ref{fig:num_sk}~(a).
Figure~\ref{fig:num_sk}~(b) presents the number of signals integrated over the last 12~hours for different distances.
The expected background counts for low, medium, and high reactor activities in SK are 4.6, 6.2 and 8.1~events. 
Although the background rate is much higher than that in KamLAND, the large target volume allows SK to collect signal events an order of magnitude larger than KamLAND.
Thus the statistical significance in SK can increase rapidly when approaching core collapse.

\begin{figure}
\centering
\includegraphics[width=0.66\textwidth]{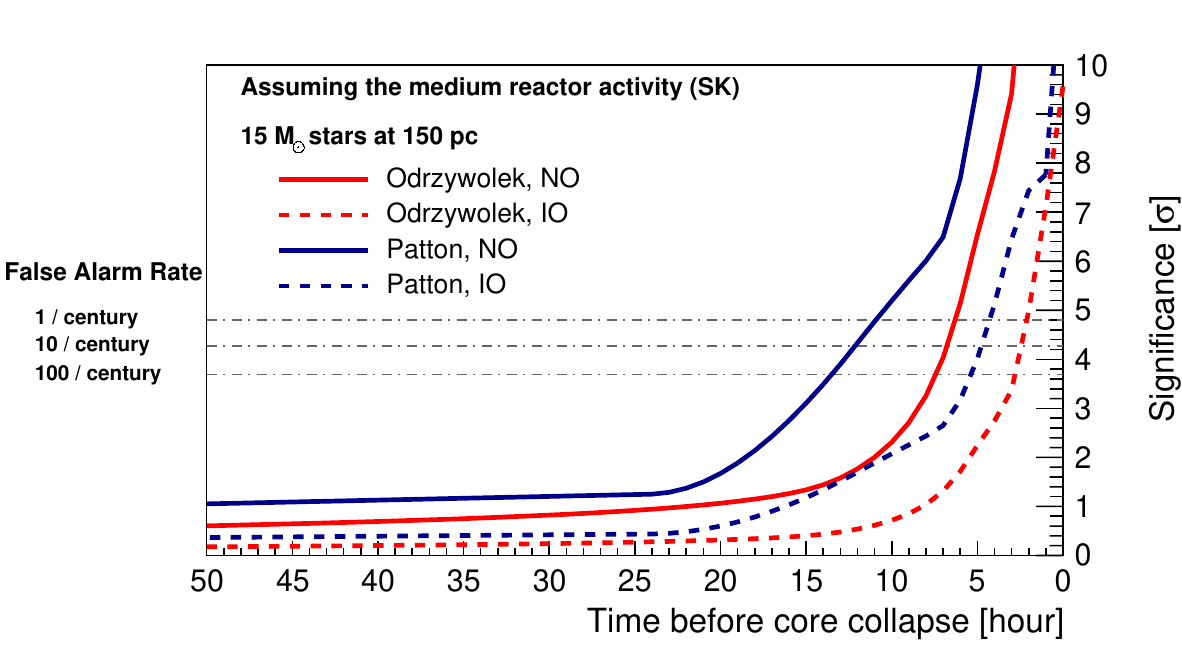}
\caption{\label{fig:sig_sk} Time evolution of the sensitivity to pre-SN neutrinos in SK with 0.03$\%$ Gd concentration, assuming medium reactor activity, following the Odrzywolek model (red) and the Patton model (blue). Solid (dashed) lines are for normal (inverted) neutrino mass ordering. Horizontal dotted-dashed lines indicate false alarm rates of 1, 10, and 100~per century. }
\end{figure}

Based on the estimations of signal and background, we assessed the sensitivity of pre-SN neutrino detection in SK-Gd with $0.03\%$ Gd loading.
Figure~\ref{fig:sig_sk} presents the time evolution of the expected detection significance in SK-Gd, assuming medium reactor activity.
The results show that SK-Gd is capable of providing an early warning before the CCSN, at most 10.9~hours assuming the Patton model and normal ordering, and at least 2.1~hours for the Odrzywolek model and inverted ordering.
The warning time for all of the simulated scenarios are summarized in Table~\ref{tab:cmb_time}.
If the reactor neutrino fluxes around the Kamioka mine are doubled, the warning time can be shortened by 0.2-1.1~hours.

\begin{figure}[t]
\centering
\includegraphics[width=0.6\textwidth]{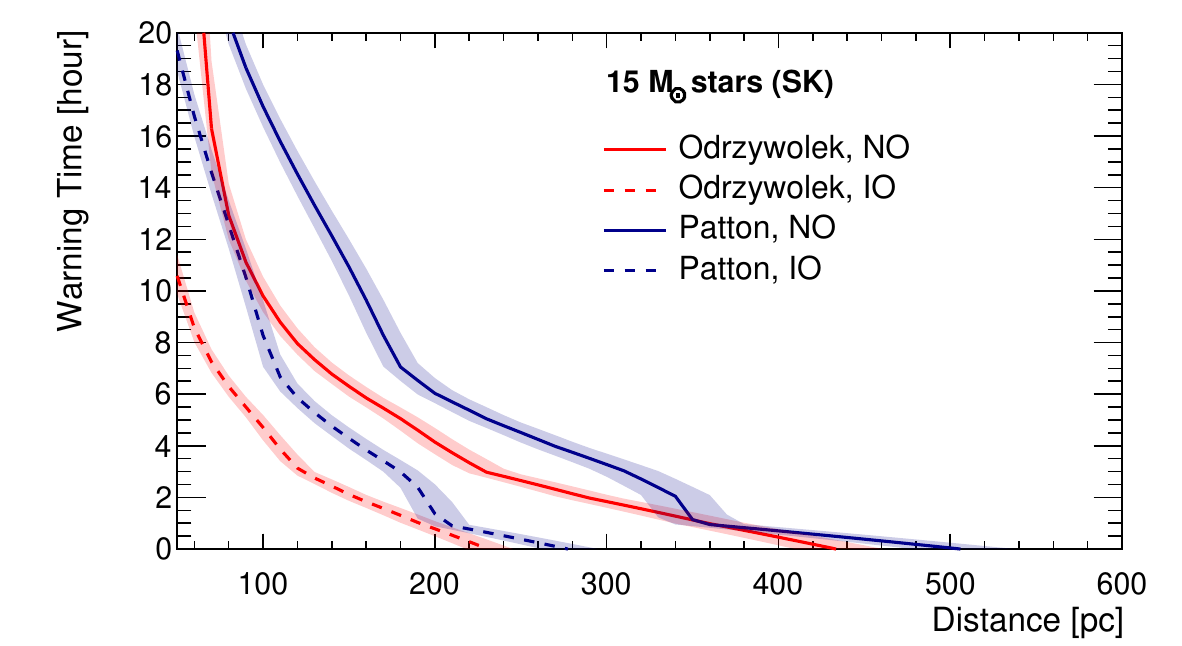}
\caption{\label{fig:dis_sk} Expected warning time in SK as a function of distance. The lines are estimations assuming medium reactor activity. The upper (lower) edges of the bands are for the low (high) reactor activity case. }
\end{figure}

We plot the warning time as a function of distance in Figure~\ref{fig:dis_sk}.
The upper (lower) edges of the bands are for low (high) reactor activity, and the lines in between are for medium reactor activity.
Under neutrino flux assumptions of the Patton model and normal ordering, the SK alert can cover 15\solarmass stars with a distance of 500\parsec from Earth, for the medium reactor activity case.


\subsection{Discussion} \label{subsec:discussion}
%


The results shown above suggest that the two detectors, KamLAND and SK, have different advantages in pre-SN neutrino detection.
For KamLAND, the low background rate is an advantage in terms of resolving a small signal.
As shown in Figure~\ref{fig:num_kl} (a), for a Betelgeuse-like pre-SN candidate, the expected number of signal events exceed the background count even tens of hours prior to the CCSN.
As a result, KamLAND can provide a warning for nearby supernovae earlier than SK.
On the other hand, the number of signal events in KamLAND is limited by the target mass, making it hard to reach stars at far distances.
SK is sensitive to CCSN candidates further away from Earth compared to KamLAND.
The large target mass of SK can increase the significance rapidly when approaching the CCSN, resulting in a much higher $\bar{\nu}_e$ event rate.
But due to its relatively high background rate, SK is less sensitive to small signals.
By noting that these two detectors are complementary in pre-SN neutrino searches, a joint search combining measurements from these two detectors should improve the current detection sensitivity.
We show in Section~\ref{sec:cmbsen} that the combined alert benefits from the advantages of these two detectors.
We expect the complementary properties of the two detectors to create a synergistic bond, enhancing the sensitivity to pre-SN neutrino signals.

\section{Combined sensitivity to pre-SN neutrinos} \label{sec:cmbsen}
\subsection{Statistical approach for the combined search} \label{subsec:cmb_fit}

The purpose of the combined pre-SN alert system is to provide a semi-realtime result of an online search of pre-SN neutrino signals. 
The chosen strategy is to perform a test of significance every five minutes upon the observed numbers of candidates and the expected numbers of background events in KamLAND and SK.
The following likelihood function is constructed, which is a product of the Poisson likelihood of each experiment,

\begin{equation}
    \mathcal{L}_{\mathrm{combine}} = \mathcal{L}_{\mathrm{SK}} \times \mathcal{L}_{\mathrm{KL}}.
\end{equation}
The test statistic based on likelihood ratio $\Lambda_{\mathrm{combine}}$ can be calculated by substituting $\mathcal{L}_{\mathrm{combine}}$ for $\mathcal{L}_{\mathrm{x}}$ in Equation~\ref{eq:likelihoodratio}.
The corresponding significance is found by calculating the chi-square quantile for two degrees of freedom.

For any pre-SN neutrino model, the neutrino fluxes in both detectors should be the same, and thus there are correlations between numbers of signals in KamLAND and SK. 
However, we note that the test statistic $\mathcal{L}_{\mathrm{combine}}$ merely reflects the level of agreement between data and the background-only hypothesis.
Whether the data is consistent with a pre-SN neutrino model is not tested.
Therefore, the parameters of signal in SK and KamLAND ($S_{\mathrm{SK}}$ and $S_{\mathrm{KL}}$) are treated as two independent parameters, and the degrees of freedom is considered as two.

Likewise, reactor neutrino background in the two detectors is correlated.
This correlation does not affect the statistical test, because the expected number of background $B$ is estimated from the data taken in the background time window and normalized to the analysis time window, based on the assumption that the background rate is consistent with the background data taken before. 
However, such an assumption may not hold, since the reactor $\bar{\nu}_e$ background can change significantly within a week if nuclear reactors are turned on or off.
An unexpected increase of reactor $\bar{\nu}_e$ flux, for example, when several nearby reactors are turned on, may cause an excess of $\bar{\nu}_e$ events in the detectors.
Although the background rate is still far below the alert criteria even in an extreme case we can imagine, i.e. the high reactor activity scenario, a higher baseline can increase the risk of sending a false alarm.
The reactor neutrino background is irreducible, as it consists of true $\bar{\nu}_e$ events and its energy range overlaps that of the signal.
Concerning this issue, we perform frequent background measurements in both detectors.


\begin{figure}[h]
\centering
\includegraphics[width=0.6\textwidth]{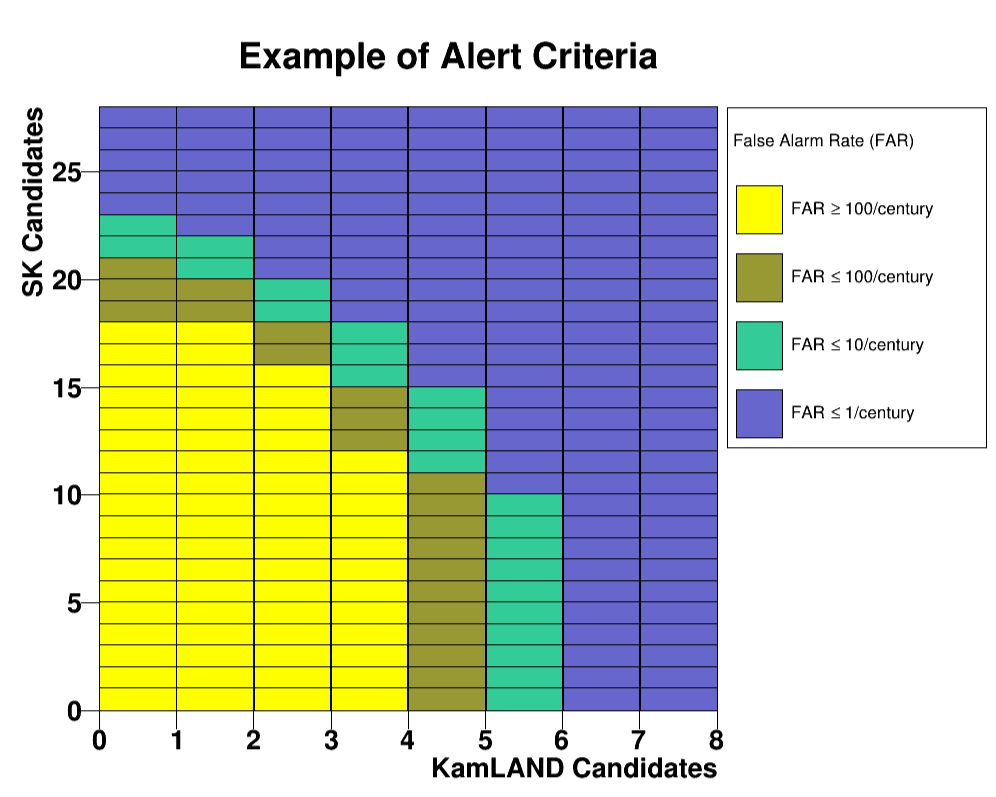}
\caption{\label{fig:map_null} Example contour of the false alarm rates extracted from toy Monte-Carlo simulation assuming the background-only hypothesis. The x-axis and y-axis are numbers of candidates observed in KamLAND (in 24 hours) and SK (in 12 hours). The assumed background rates are 12.4~events per day in SK and 0.19~events per day in KamLAND, the same as those in the medium reactor activity scenario. }
\end{figure}

As mentioned in Section~\ref{subsec:far}, the frequency to find a single combined search at least as extreme as the observation, i.e. the false alarm rate, is utilized to report the results of the pre-SN neutrino search.
Figure~\ref{fig:map_null} shows an example contour of the false alarm rate resulting from toy Monte-Carlo simulation. 
The x-axis and y-axis are numbers of candidates observed in KamLAND (in 24 hours) and SK (in 12 hours), respectively. 
The color of each box indicates the false alarm rate level of the corresponding numbers of observed events. 
If the observation drops in the yellow regions, which indicates a ${\text{false alarm rate} \geq 100~\text{per century}}$, there is no hint of a pre-SN neutrino emission. 
If the observation is in the blue regions, where the ${\text{false alarm rate} \leq 1~\text{per century}}$, it is considered as a significant excursion of the observed event rate which could be caused by pre-SN $\bar{\nu}_e$.

\subsection{Combined sensitivity}

\begin{figure}[htb!]
    \centering
    \includegraphics[width=0.66\textwidth]{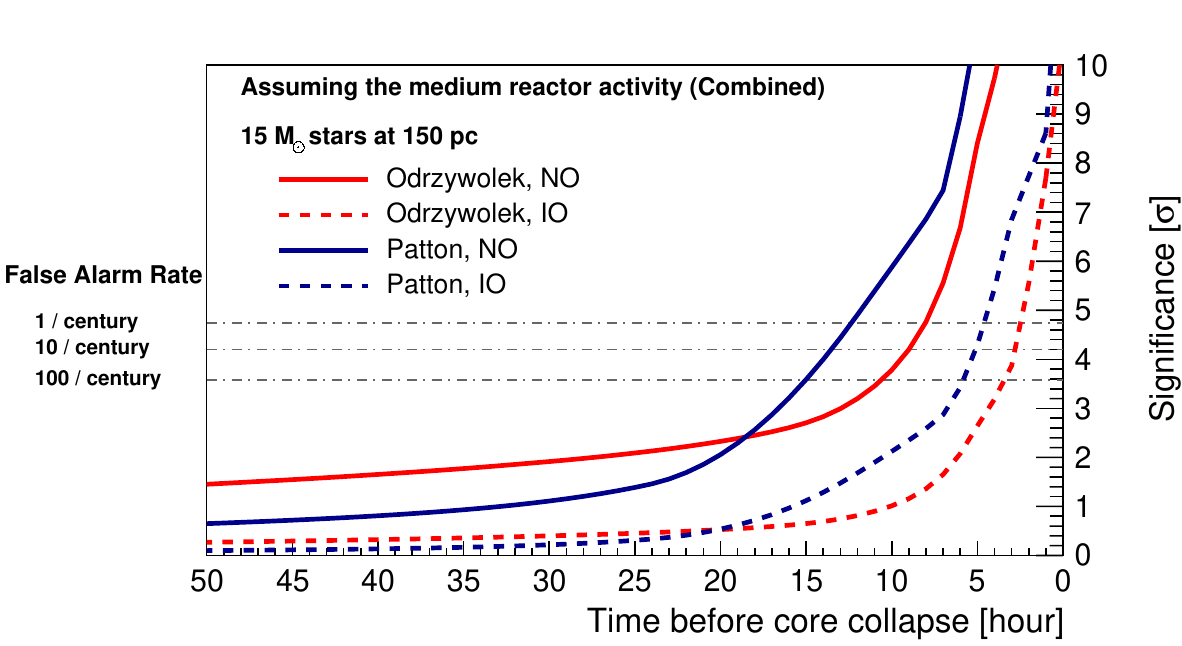}
    \caption{Combined sensitivity to pre-SN neutrinos as a function of time based on the detection capability of KamLAND and SK-Gd with 0.03$\%$ Gd concentration, assuming medium reactor activity, following the Odrzywolek model (red) and the Patton model (blue). Solid (dashed) lines are for normal (inverted) neutrino mass ordering. Horizontal dotted-dashed lines indicate false alarm rates of 1, 10, and 100~per century.}
    \label{fig:cmb_sig}
\end{figure}

Following the above mentioned statistical approach, we perform a joint sensitivity study based on the estimations of $S_{\mathrm{KL}}$, $S_{\mathrm{SK}}$, $B_{\mathrm{KL}}$ and $B_{\mathrm{SK}}$ presented in Section~\ref{sec:indsen}.
Figure~\ref{fig:cmb_sig} shows the time evolution of the combined sensitivity of pre-SN neutrinos from a Betelgeuse-like star of 15\,M$_{\odot}$, assuming medium reactor activity.
For the Patton model and normal mass ordering, the warning time is extended to 12.4~hours prior to the CCSN.
Similar to the discussions in Section~\ref{sec:indsen}, we assessed the sensitivities for the two pre-SN models, the three reactor activity cases, and the two neutrino mass orderings.
Table~\ref{tab:cmb_time} summarizes the warning time of the combined alert and the individual alerts.
These results indicate that the combined alert presents an improved performance, because the warning times are longer compared to either of the individual alerts.
It is important to note that, even in the high reactor activity case, the warning time is at least 2.2~hours before the CCSN.

\begin{deluxetable}{ccc|ccc} \label{tab:cmb_time}
\tablecaption{Warning time of the KamLAND-only, SK-only, and combined search for each pre-SN neutrino model, neutrino mass ordering and reactor activity, assuming a Betelgeuse-like pre-SN star of 15\,M$_{\odot}$. The latency due to data processing is not taken into account.}
\tablewidth{0pt}
\tablehead{
\colhead{} & \colhead{} & \colhead{} & \multicolumn{3}{c}{Warning time [hour]} \\
\colhead{Alert system} & \colhead{Pre-SN model} & \colhead{Mass ordering} & \colhead{Low reactor activity} & \colhead{Medium reactor activity} & \colhead{High reactor activity} \\
}
\startdata
KamLAND &   Odrzywolek   & NO & 8.3 & 6.5 & 5.5 \\
        &                & IO & 0.9  & N/A  & N/A \\
        &   Patton       & NO & 8.1 & 6.1 & 5.0 \\
        &                & IO & 0.8  & 0.2  & N/A \\
\hline
SK      &   Odrzywolek   & NO & 6.7 & 6.3 & 5.9 \\
        &                & IO & 2.4  & 2.1  & 1.9 \\
        &   Patton       & NO & 12.0 & 10.9 & 9.8 \\
        &                & IO & 4.7  & 4.3  & 3.9\\
\hline
Combined &  Odrzywolek   & NO & 9.8 & 8.0 & 7.3 \\
        &                & IO & 3.0  & 2.5  & 2.2 \\
        &   Patton       & NO & 14.2 & 12.4 & 11.2 \\
        &                & IO & 5.4  & 4.6  & 4.2 \\
\enddata
\tablecomments{N/A denotes not applicable, meaning the expected significance does not reach the alert criteria.}
\end{deluxetable}

\begin{figure}[htb!]
    \centering
    \includegraphics[width=0.6\textwidth]{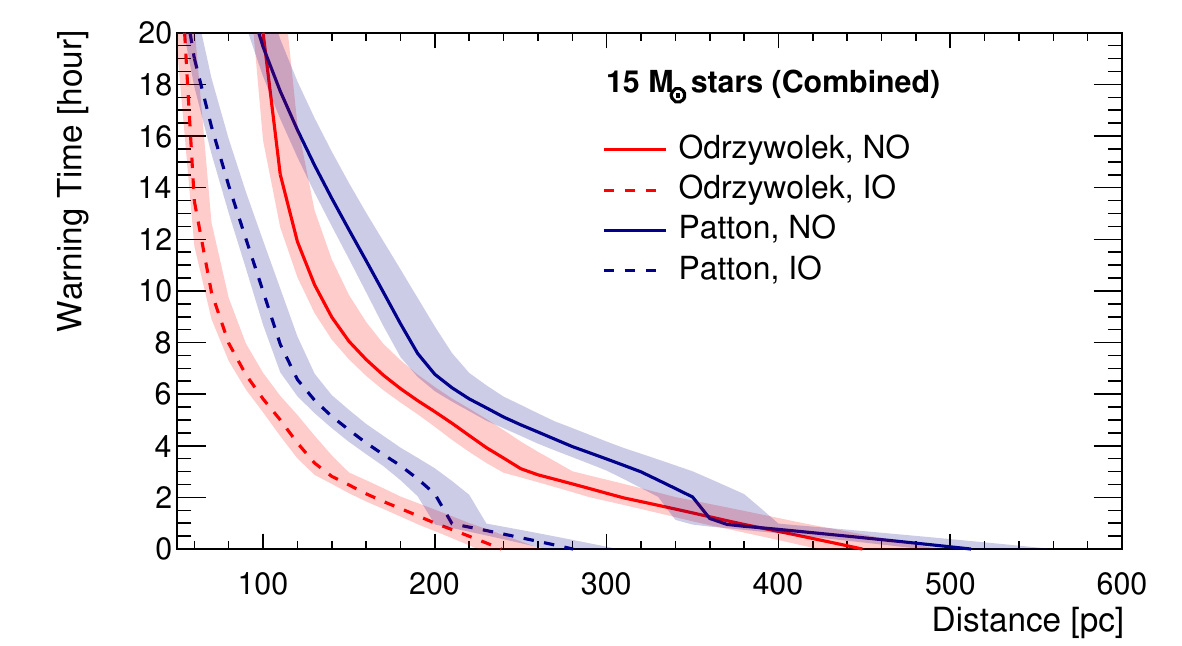}
    \caption{ Expected warning time of the combined search as a function of distance. The lines are estimations assuming medium reactor activity. The upper (lower) edges of the bands are for the low (high) reactor activity case. }
    \label{fig:cmb_whdis}
\end{figure}

Figure~\ref{fig:cmb_whdis} presents the expected warning time and the star distance coverage of the combined alert.
Variations due to changes in the reactor neutrino flux are shown as shaded, enveloped by the upper edges resulting from low reactor activity, and the lower edges resulting from high reactor activity.
Significant improvements in star distance coverage are observed when comparing to the individual alerts shown in Figure~\ref{fig:dis_kl} and Figure~\ref{fig:dis_sk}.
Assuming 15\solarmass stars, the combined alert is able to cover 510\parsec for the medium reactor activity case.

These results demonstrate the complementarity of the KamLAND and the SK-Gd detectors.
Taking advantage of the low background rate of KamLAND and the large target mass of SK, the combined alert presents improvement in extending the warning time as well as the distance coverage.
In light of this, a combined pre-SN neutrino alert system was developed, and will be discussed in Section~\ref{sec:online}.

\section{Combined online search for pre-SN neutrinos} \label{sec:online}

The combined pre-SN alert system aims to provide early warning of a potential CCSN upon the detection of pre-SN neutrinos in the KamLAND and SK detectors.
It is now operational, ready to issue alarms of CCSNs. 

\begin{figure}[htb!]
    \centering
    \includegraphics[width=0.8\textwidth]{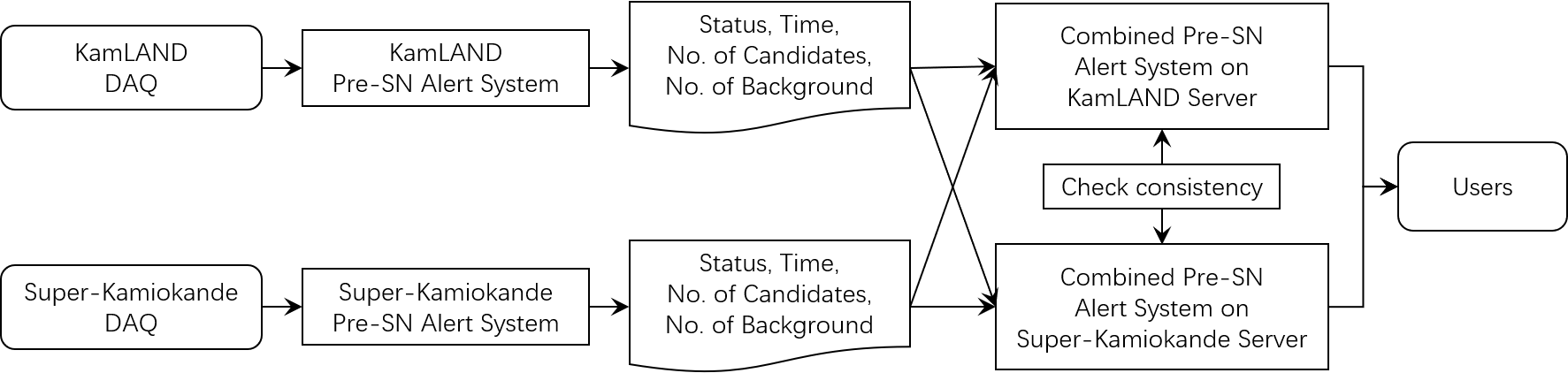}
    \caption{Illustration of the input of the combined pre-SN alert system.}
    \label{fig:dataflow}
\end{figure}

The workflow of the system is introduced in the following.
The system receives from both detectors the number of pre-SN neutrino candidates and the expected number of background.
Processed by the DAQ systems, events in the two detectors are selected by their own selection processes following the descriptions in Section~\ref{subsec:klsel} and Section~\ref{subsec:sksel}.
The individual pre-SN alert software of KamLAND (SK) then counts the number of observed candidates $N_{\mathrm{KL}}$ ($N_{\mathrm{SK}}$) within a 24-hour (12-hour) time window.
The expected number of background $B_{\mathrm{KL}}$ ($B_{\mathrm{SK}}$) is estimated using data from a background time window of $\sim$90 ($\sim$30) days, as described in Section~\ref{subsec:anastg}.
Validity of data is also taken into consideration. 
Detector status is monitored and marked by a status code. 
When a detector undergoes calibration work, a test run or shutdown, the status of the detector is marked as ``abnormal" and the data from this period will be invalidated.
In addition to the above situations, there may also be network connection problems which can delay the data transfer.
Therefore, the differences between the current time and the time when data are processed will also be checked.
The individual KamLAND and SK pre-SN alert systems gather the above information, and exchange them between the servers of KamLAND and SK, as illustrated in Figure~\ref{fig:dataflow}.
The update frequency of the input is once every 5~minutes for each of the experiments. 
These inputs will be processed by the combined pre-SN alert software, yielding a result of the combined pre-SN search.
The result will be exported to users, and the alert decision will be made based on the result.

\begin{figure}[htb!]
    \centering
    \includegraphics[width=1.0\textwidth]{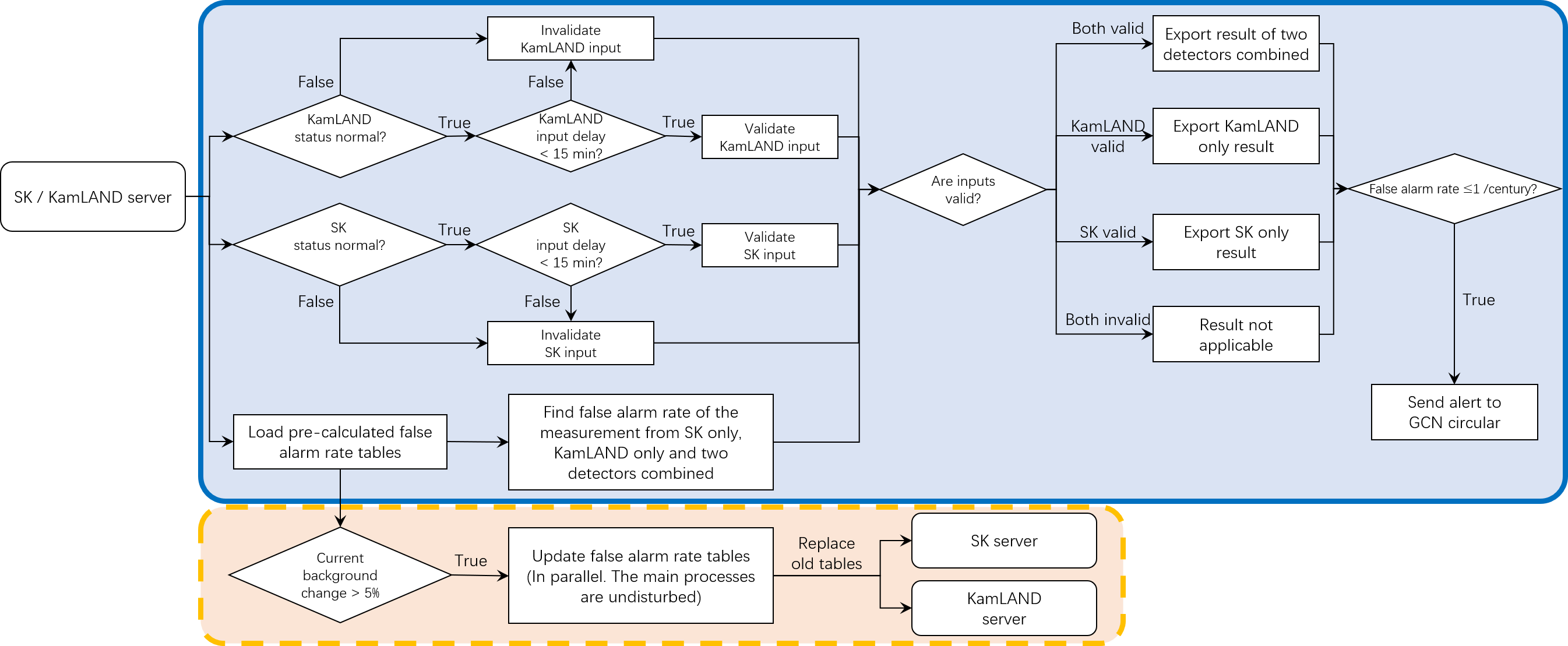}
    \caption{Illustration of the workflow of the combined pre-SN alert system. The main processes to produce results of statistical tests as pictured in the blue box with solid border are identical for the systems on both servers. The process to update false alarm rate tables as plotted in the orange box with dashed border is installed exclusively on the SK server. }
    \label{fig:workflow}
\end{figure}

On the two servers, identically functioning software for the combined pre-SN alert system are installed.
If either one of the software pipelines fails, the other one can still output the search result and issue alerts.
Figure~\ref{fig:workflow} illustrates the workflow of this software.
The software runs on a precise repeating schedule once every 5~minutes.
Validity of inputs is first checked by examining the status of each detector and the timestamps of the inputs.
If a detector has an abnormal status, or if data from it is delayed for over 15~minutes, input from this detector is invalidated.
In this case, instead of exporting the combined search result, the system will output a result based only on the valid input.
If all of the inputs are invalid, the result is not applicable.
Only when both KamLAND and SK are in normal status and the data are up-to-date, the system exports the result of the combined search.

In order to determine the corresponding false alarm rate, the software loads three pre-calculated false alarm rate tables, for KamLAND-only, SK-only, and their combination.
Once the software finds a ${\text{false alarm rate} \leq 1~\text{per century}}$, an alarm will be sent to the Gamma-ray Coordinates Network (GCN) via an email-based circular.
In addition, a text file containing the false alarm rate, along with a timestamp, and a code that denotes whether the result is for KamLAND-only, SK-only or the combined search, is available to users who have registered on the official website of the combined pre-SN alert system (\url{https://www.lowbg.org/presnalarm/}).

The above processes, called main processes, are identical in both servers, as pictured in the blue box with solid border in Figure~\ref{fig:workflow}.
An additional process as shown in the orange box with dashed border is uniquely installed on the SK server. 
In this process, the false alarm rate tables are updated automatically upon any changes $>5\%$ in the expected numbers of background. 
This process, typically takes $\sim$40~minutes, and is in parallel with the main processes, in order not to delay the output of the results. 
Therefore, the pre-calculated false alarm rate tables do not always correspond to the current background values.
However, we note that a significant change in the background rates within an hour is unusual, because the background rates are obtained from measurements of a specific time window long enough to mitigate the effects of statistical fluctuations.

This alert system, operational since May 2023 and accessible to the public, is designed to promptly notify astronomers and particle physicists to maintain operational readiness at their observatories, ensuring they do not miss any impending supernova events.
Users can acquire identical results contained in the above mentioned text file from either of the servers, and are encouraged to check the consistency of the results from the two servers before putting into scientific use.
Directional information of pre-SN neutrinos is not available from either SK or KamLAND.
More information can be found on the official website (\url{https://www.lowbg.org/presnalarm/}).

\section{Conclusion} \label{sec:conc}


In this study, we present updates on the sensitivity to pre-supernova neutrinos from a Betelgeuse-like star of the individual alert systems of KamLAND and Super-Kamiokande, and introduce a combined pre-supernova alert system with the two detectors. 
Pre-supernova neutrino fluxes are calculated based on the models from \citep{odrzywolek} and \citep{patton}, with neutrino oscillation effects.
Due to their similar energy range, reactor neutrinos originating from Japanese nuclear reactors constitute an important source of background for pre-supernova neutrinos. 
Different reactor activity conditions in Japan are considered in this study, where reactor fluxes vary from low to high.
The following results are estimated assuming that nuclear reactors near the Kamioka mine (Mihama 3 unit, Ohi 3, 4 units and Takahama 1, 2, 3, 4 units) operate with a 100$\%$ load factor.
The corresponding background rates are 0.19 events per day and 12.4 events per day in KamLAND and Super-Kamiokande, respectively.

The best warning times are attained by each detector under different neutrino flux assumptions.
In the ideal case, for the Odrzywolek model and normal ordering, KamLAND can provide an early warning 6.5~hours prior to core collapse, and a pre-supernova neutrino emission can be observed up to 280\parsec from Earth.
The Super-Kamiokande pre-supernova alert has an optimistic warning time of 10.9~hours, and is able to cover a distance of 500\,pc, assuming the Patton model and normal ordering.

The combined pre-supernova alert system performs a joint statistics test based on the data from the KamLAND and the Super-Kamiokande detectors.
It has been operational and accessible to the public since May 2023. 
Integrating the complementary properties of the two detectors, the combined alert shows improved sensitivity to pre-supernova neutrinos.
An optimistic warning time of 12.4~hours is obtained, for the Patton model and normal ordering, 1.5~hours longer than the Super-Kamiokande alert and $\sim$6.3~hours longer than the KamLAND alert, with the medium reactor activity assumption.
At the same background level, its distance coverage for 15\solarmass progenitors is 510\,pc, which exceeds those of the individual alerts.
While doubling the neutrino fluxes from nearby reactors increases significantly the backgrounds and affects the sensitivity of both detectors individually, the combined alert remains sensitive to pre-supernova neutrino emission with an expected warning time of no less than 2.2~hours for a Betelgeuse-like pre-supernova star, sufficiently long to cover the latency due to data processing. 
In addition, the combined alert system reduces the dead time for pre-supernova neutrino detection, promoting continuous monitoring even if one of the detectors is temporarily offline.
All of these demonstrate the benefits of having a combined search for pre-supernova neutrinos.

\begin{acknowledgements}
\section*{Acknowledgements}
The KamLAND collaboration and the SK collaboration gratefully acknowledge the cooperation of the Kamioka Mining and Smelting Company.
KamLAND is supported by MEXT KAKENHI Grant Numbers 19H05802;  the World Premier International Research Center Initiative (WPI Initiative), MEXT, Japan; Netherlands Organization for Scientific Research (NWO); and under the U.S. Department of Energy (DOE) Contract No. DE-AC02-05CH11231, the National Science Foundation (NSF) No. NSF-1806440, NSF-2012964, as well as other DOE and NSF grants to individual institutions.
The KamLAND collaboration thank the support of NII for SINET.
The SK experiment has been built and operated from funding by the Japanese Ministry of Education, Culture, Sports, Science and Technology, the U.S. Department of Energy, and the U.S. National Science Foundation. 
Some of the SK collaborators have been supported by funds from the National Research Foundation of Korea (NRF-2009-0083526 and NRF 2022R1A5A1030700) funded by the Ministry of Science, Information and Communication Technology (ICT); the Institute for Basic Science (IBS-R016-Y2); 
and the Ministry of Education (2018R1D1A1B07049158, 2021R1I1A1A01042256, 2021R1I1A1A01059559); 
the Japan Society for the Promotion of Science; the National Natural Science Foundation of China under Grants No.11620101004; 
the Spanish Ministry of Science, Universities and Innovation (grant PID2021-124050NB-C31); 
the Natural Sciences and Engineering Research Council (NSERC) of Canada; the Scinet and Westgrid consortia of Compute Canada; 
the National Science Centre (UMO-2018/30/E/ST2/00441 and UMO-2022/46/E/ST2/00336) and the Ministry of Science and Higher Education (2023/WK/04), Poland; 
the Science and Technology Facilities Council (STFC) and Grid for Particle Physics (GridPP), UK; 
the European Union’s Horizon 2020 Research and Innovation Programme under the Marie Sklodowska-Curie grant agreement no. 754496; H2020-MSCARISE-2018 JENNIFER2 grant agreement no.822070, H2020-MSCA-RISE-2019 SK2HK grant agreement no. 872549; and European Union's Next Generation EU/PRTR grant CA3/RSUE2021-00559.
\end{acknowledgements}

\bibliography{main}{}
\bibliographystyle{aasjournal}

\allauthors

\end{document}